\documentclass[letterpaper]{article} 
\usepackage{aaai23}  
\usepackage{times}  
\usepackage{helvet}  
\usepackage{courier}  
\usepackage[hyphens]{url}  
\usepackage{graphicx} 
\usepackage{natbib}  
\usepackage{caption} 
\usepackage{algorithm}
\usepackage{algorithmic}
\usepackage{times}
\usepackage{soul}
\usepackage{url}
\usepackage[hidelinks]{hyperref}
\usepackage[utf8]{inputenc}
\usepackage{caption}
\usepackage{graphicx}
\usepackage{amsmath}
\usepackage{amsfonts}
\usepackage{amsthm}
\usepackage{booktabs}
\usepackage{bm}
\usepackage{multirow}
\usepackage{float}
\usepackage{tabularx,ragged2e}
\usepackage{mathtools}
\usepackage{subcaption}
\usepackage{pifont}
\usepackage[dvipsnames]{xcolor}
\DeclareMathOperator*{\argmin}{arg\,min}
\usepackage{newfloat}
\usepackage{listings}
\DeclareCaptionStyle{ruled}{labelfont=normalfont,labelsep=colon,strut=off} 
\floatstyle{ruled}
\newfloat{listing}{tb}{lst}{}
\floatname{listing}{Listing}
\title{REGAS: REspiratory-GAted Synthesis of Views for Multi-Phase CBCT Reconstruction from a single 3D CBCT Acquisition}
\author {
    Cheng Peng\textsuperscript{\rm 1}, 
    Haofu Liao\textsuperscript{\rm 2},
    S. Kevin Zhou\textsuperscript{\rm 3,4},
    Rama Chellappa\textsuperscript{\rm 1}
}
\affiliations {
    \textsuperscript{\rm 1} Johns Hopkins University, 
    \textsuperscript{\rm 2} University of Rochester\\
    \textsuperscript{\rm 3} School of Biomedical Engineering \& Suzhou Institute for Advanced Research
    Center for Medical Imaging, Robotics, and Analytic Computing \& LEarning
    (MIRACLE), University of Science and Technology of China\\
    \textsuperscript{\rm 4} Key Lab of Intelligent Information Processing of Chinese Academy of Sciences (CAS),
    Institute of Computing Technology\\
    cpeng26@jh.edu
}

\usepackage{bibentry}

\begin{document}

\maketitle

\begin{abstract}
It is a long-standing challenge to reconstruct Cone Beam Computed Tomography (CBCT) of the lung under respiratory motion. This work takes a step further to address a challenging setting in reconstructing a \textbf{multi-phase} 4D lung image from just a \textbf{single} 3D CBCT acquisition.
To this end, we introduce \underline{RE}piratory-\underline{GA}ted \underline{S}ynthesis of views, or \textbf{REGAS}. REGAS proposes a self-supervised method to synthesize the undersampled tomographic views and mitigate aliasing artifacts in reconstructed images. This method allows a much better estimation of between-phase Deformation Vector Fields (DVFs), which are used to enhance reconstruction quality from direct observations \textit{without synthesis}. To address the large memory cost of deep neural networks on high resolution 4D data, REGAS introduces a novel Ray Path Transformation (RPT) that allows for \textit{distributed, differentiable} forward projections. REGAS require no additional measurements like prior scans, air-flow volume, or breathing velocity. Our extensive experiments show that REGAS significantly outperforms comparable methods in quantitative metrics and visual quality.

\end{abstract}
\section{Introduction}
\begin{figure}[!tb]
    \setlength{\abovecaptionskip}{3pt}
    \setlength{\tabcolsep}{2pt}
    \vspace{-1em}
    \begin{tabular}[b]{cc}
        \multicolumn{2}{c}{
        \begin{subfigure}[b]{\linewidth}
            \includegraphics[width=\textwidth]{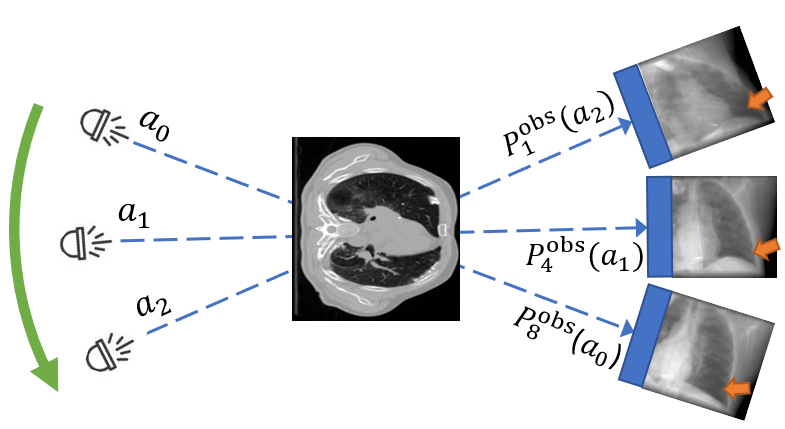}
            \caption{CBCT acquisitions}
            \label{CBCT_acq}
        \end{subfigure}
        } \\
        \begin{subfigure}[b]{.245\linewidth}
            \includegraphics[width=\textwidth]{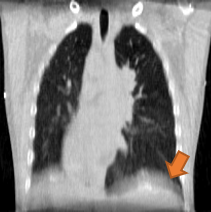}
            \caption{FDK}
            \label{FBP}
        \end{subfigure} &
        \begin{subfigure}[b]{.755\linewidth}
            \includegraphics[width=\textwidth]{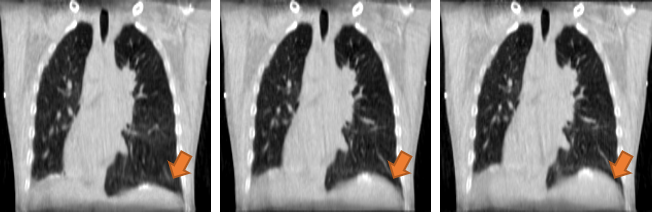}
            \caption{Ours, shown in phase 1, 3, and 5}
            \label{RGPS}
        \end{subfigure} \\
    \end{tabular}
    \caption{The respiratory motion leads to inconsistencies between projections, thus the traditionally reconstructed image, \textit{e.g.} in \ref{FBP}, contains significant motion blur. Our proposed REGAS can disentangle the observed motion, as indicated by the orange arrows in \ref{RGPS}.}
    \label{fig:intro}
    \vspace{-1.5em}
\end{figure}

Cone Beam Computed Tomography (CBCT) is a medical imaging technique widely adopted in dentistry, image-guided radiation therapy, etc. CBCT acquires 2D tomographic views of 3D object by emitting radiation from a rotating gantry, as shown in Fig.~\ref{CBCT_acq}, and use these views for 3D reconstruction~\cite{Feldkamp:84}. Compared to traditional CT, a CBCT scanner emits less radiation on patients~\cite{doi:10.3109/0284186X.2011.590525}, is less expensive, and is more flexible. 


   
   Accurate lung CBCT imaging is of significant clinical value, especially given the Coronavirus pandemic. However, lung CBCT imaging is affected by the \textit{motion-induced} inconsistencies between acquired views and the subsequent motion artifacts in reconstructions, as shown in Fig.~\ref{FBP}. 
   A common way to address this is to model the breathing motion as a cyclical process and approximate it by a number of 3D CBCTs, referred to as \textit{phases}. Under this model, each view is attributed to one of the phases for reconstruction. The phase-view correspondence can be obtained either through dedicated physical devices~\cite{Beaudry_2015} or through signal processing on acquisition data~\cite{yan2013extracting}. This is known as respiratory-gated 4D CBCT reconstruction. While motion is disentangled by the phase dimension, each phase corresponds to only a portion of the acquired views; therefore, \textit{a high quality 4D CBCT requires a large number of views to faithfully reconstruct each phase}. This leads to a trade-off between a longer scan time and higher radiation dosage, or lower image qualities with few views, similar to a regular 3D CBCT acquisition.

   
   Reconstructing 4D CBCT with limited acquisitions is a difficult problem. Existing methods (\romannumeral 1) rely on handcrafted regularizers to address the aliasing artifacts, e.g. through spatial or temporal Total Variation (TV)~\cite{andersen1984simultaneous,blondiaux2017imagerie}, or (\romannumeral 2) require additional information like prior scans, air-flow volume and velocity, deformation vector field (DVF) model, etc.~\cite{brock2010reconstruction,chee2019mcsart,ren2012development,wang2013simultaneous}, which introduces additional devices and costs into the process. 
   Many deep learning based (DL-based) methods use supervised learning to suppress sparse-view artifacts, e.g. \cite{pmid32575088,DBLP:journals/tmi/HanY18,https://doi.org/10.1002/mp.15183}; however, these end-to-end methods are applicable only to 2D CT due to the heavy computational and data requirements. Furthermore, factors such as scanner differences and breathing motion differences can lead to generalization issues from supervised training.

We propose a novel 4D CBCT reconstruction method called \underline{RE}spiratory-\underline{GA}ted \underline{S}ynthesis of views (\textbf{REGAS}), which combines the advantages of traditional methods and DL-based methods. We note that a key component to address 4D CBCT reconstruction is inter-phase deformation modeling; i.e., we can obtain high quality 4D CBCT reconstruction if the breathing DVF is known. However, obtaining such DVF is extremely difficult with limited acquisitions due to noisy reconstruction and a lack of groundtruth. To address the reconstruction, REGAS first synthesizes the unobserved views of each phase through a Convolutional Neural Network (CNN). The synthesized views are then combined with the observed views for reconstruction, which addresses the undersampled condition and the subsequent streaky artifacts. More accurate DVFs are estimated based on our improved reconstructions, and are used to further update individual phases from observed views only. 

REGAS has several advantages compared to previous DL-based methods. Firstly, REGAS is \textbf{self-supervised}; it synthesizes unobserved views by leveraging the
3D geometric consistency from undersampled phases and the texture similarity to observed views, which is captured by a self-supervised adversarial loss. Furthermore, REGAS is \textbf{scalable to high resolution} through our novel Ray Path Transformation (RPT), which allows for distributed, differentiable forward projections. Therefore, REGAS can synthesize views that are \textbf{geometrically consistent and of high fidelity}, while requiring no groundtruth for training. The synthesized views lead to higher quality reconstructions, which assist in better DVF estimation. The accurate DVF estimation enables us to obtain synthesis-free reconstructions and eliminate concerns about the reliability of generative methods.

   
   Our contributions can be summarized in four parts:

   
   \begin{enumerate}
   \item We propose REGAS, which formulates respiratory-gated 4D CBCT reconstruction as a view completion problem. REGAS incorporates the physical imaging geometry with a combination of 3D and 2D CNNs.
    \item We propose a view-dependent, ray-tracing transformation scheme called RPT, which efficiently represents high resolution 3D information relating to a view and can be easily decomposed into distributed learning.
    \item We leverage the synthetic-view-assisted reconstructions for better inter-phase DVF estimations; this enable us to then optimize 4D CBCT without synthetic data and achieve more reliable performance.
    \item We evaluate REGAS on a large scale and realistically simulated CBCT acquisition data. Our experiments show that REGAS reliably reconstructs 4D CBCT from a single 3D CBCT acquisition and significantly outperforms comparable methods.
\end{enumerate}
\section{Related Work}

\subsection{Traditional 4D CBCT Reconstruction}

In its fundamental form, 4D CBCT reconstruction is based on the FDK algorithm~\cite{Feldkamp:84}. Under the respiratory-gated model, various methods have been proposed to improve 4D CBCT reconstruction. The algebraic approach attempts to improve reconstructions iteratively based on different aspects of image quality and can be generally described as:
\begin{align}\label{eqn:iter_recon}
\hat{V} = \argmin_V ||AV - P||^2 + R(V),
\end{align}
where $V$ is the reconstructed phase, $A$ is the the projection matrix that describes the voxel-pixel correspondence between $V$ and views $P$, and $R(V)$ is a regularization term, \textit{e.g.,} total variation (TV). The family of Algebraic Reconstruction Rechniques (ART) \cite{blondiaux2017imagerie}, such as SART and OSSART, falls under this category. Prior Image Constraint Compressive Sensing (PICCS) \cite{chen2008method} seeks to minimize the total variation and pixel-wise difference between reconstructed phase and a prior image, which can be a prior scan or the reconstruction from using all the observed views.~\cite{Ritschl_2012} propose to enforce total variation along the temporal axis. ~\cite{mory2014cardiac} propose ROOSTER, which uses various regularizations to reduce noise within a selected region of interest.

DVF-based approaches seek to use the deformation fields between phases to improve reconstructions; however, additional information is often required to find such DVFs. \cite{brock2010reconstruction,ren2012development} rely on a previously obtained high-quality 4D CT scan to compute the initial DVFs, and correct them with the currently observed views; the final DVFs are then used to deform the prior high-quality 4D scan to represent patient's current state. Combinations of algebraic and DVF-based approaches also exist, with algorithms such as SMEIR~\cite{wang2013simultaneous} and MC-SART~\cite{chee2019mcsart}. These methods obtain the initial phase reconstructions from an algebraic approach, estimate DVF based on those reconstructions, and iteratively improve both the reconstructions and the DVFs. As initial reconstructions are of low quality, there is no guarantee that DVFs and reconstructions converge simultaneously. In particular, MC-SART requires additional devices to measure diaphragm amplitude and velocity.

\subsection{Deep Learning-based CT Reconstruction}
Deep learning-based methods have obtained impressive results on reconstructing sparse-view CT images~\cite{pmid32575088,DBLP:journals/tmi/HanY18,https://doi.org/10.1002/mp.15183}. However, these methods are often designed for 2D slice reconstruction, where the computational complexity is reasonable. These methods also rely on supervised training and high quality groundtruth, which limit their application in 4D CBCT as the breathing motion is continuous and does not have an easily measurable groundtruth. 4D-AirNet~\cite{pmid32575088}, for example, performs their experiments on 2D CT slices despite motivated by 4D CBCT, as the computational cost is otherwise infeasible.

\section{REspiratory-GAted Synthesis (REGAS)}

\begin{figure*}[!htb]
    \centering
      \includegraphics[width=1\textwidth]{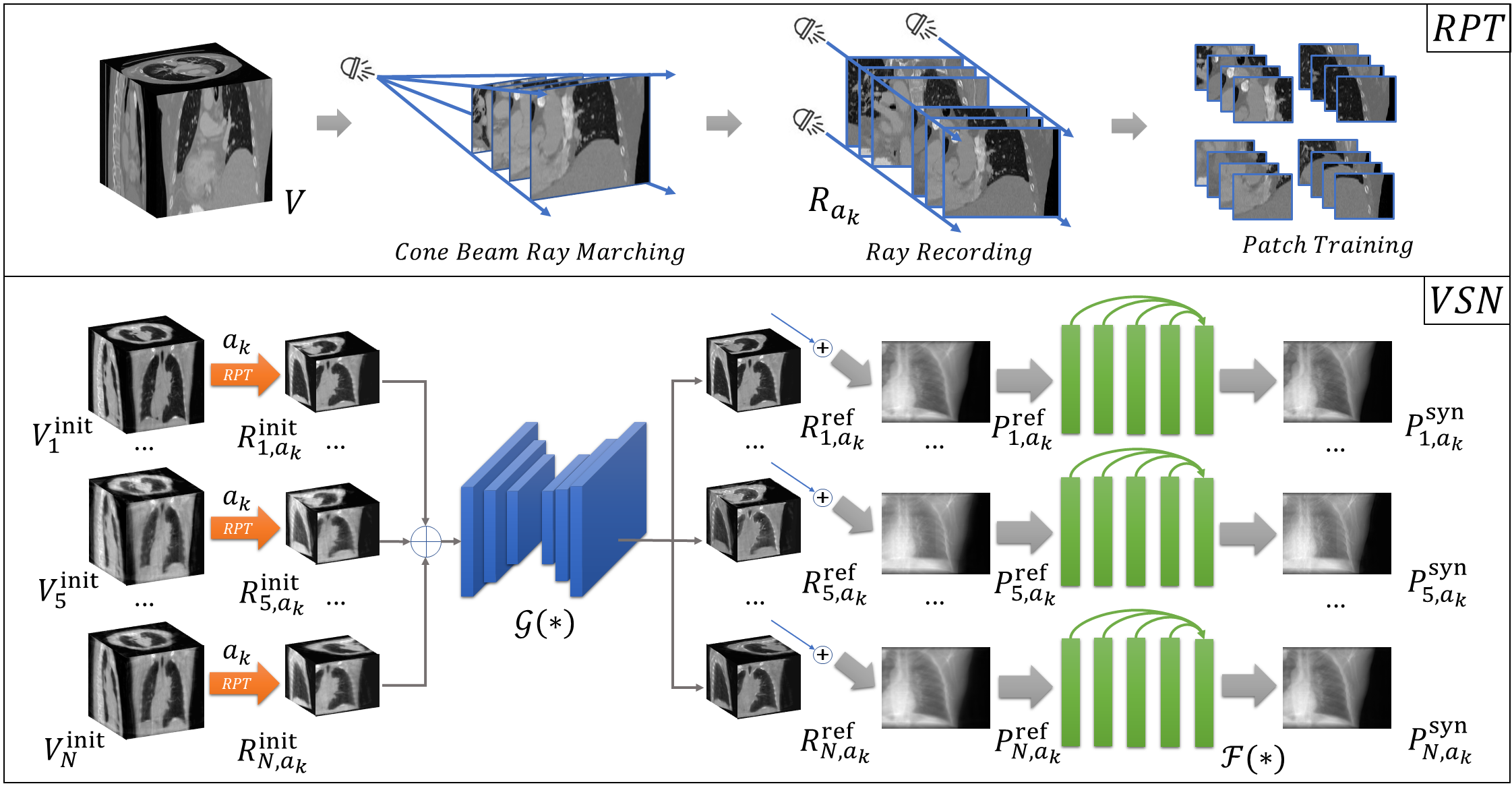}
    \caption{REGAS synthesizes unobserved views for phase $i$ based on observed views across all phases. It first takes the RPT transformed initial reconstructions $R^{\textrm{init}}_{i,a_k}$ and refines them through a 3D network $\mathcal{G}$. The resulting $R^{\textrm{ref}}_{i,a_k}$ forms views $P^{\textrm{ref}}_{i,a_k}$ and goes through a 2D network $\mathcal{F}$, yielding $P^{\textrm{syn}}_{i,a_k}$. After training, $P^{\textrm{syn}}_{i,a_k}$ are used to reconstruct $V^{\textrm{syn}}_{i}$. The RPT module generates patch separable volumes with respect to the projection angle, and is differentiable.}
    \label{fig:network}
\end{figure*}

Let $P^{\textrm{obs}} = \{P^{\textrm{obs}}(a_k)\}_{k=1}^K$ be $K$ views observed from a single CBCT acquisition process where $a_k$ denotes the angle of the $k^{\textrm{th}}$ view. Through respiratory gating, views of $P^{\textrm{obs}}$ can be binned into several phases with $P^{\textrm{obs}}=\cup_{i} P^{\textrm{obs}}_i$, where $P^{\textrm{obs}}_i$ denotes the set of views from phase $i$. After binning, we can reconstruct a CBCT image for phase $i$ using views from $P^{\textrm{obs}}_i$. However, since $|P^{\textrm{obs}}_i|\ll K$, the reconstructed image suffers from aliasing artifacts. At the core of REGAS, we propose to synthesize the unobserved views $P^{\textrm{syn}}_i$ for all $i$, such that $|P^{\textrm{obs}}_i|+|P^{\textrm{syn}}_i|=K$, and thus reconstructing CBCT images using $P^{\textrm{syn}}_i$ and $P^{\textrm{syn}}_i$ results in less aliasing artifact. In the following sections, we describe the two main components of REGAS: \emph{Ray Path Transformation} (RPT), which is a differentiable projection scheme that enables scalable view synthesis under the Cone-Beam geometry, and a \emph{View Synthesis Network} (VSN) that generates high quality $P^{\textrm{obs}}_i$ with self-supervision.

\subsection{Ray Path Transformation (RPT)}

Processing 4D CBCT data with a reasonable resolution limits the possible capacity of CNNs due to computational constraints. One way to mitigate this limitation is to reduce the input size via patch processing. Assuming the emitted rays traverse in Cartesian planes (e.g., in Parallel-Beam or Fan-Beam CT), one can separate the observed volume into slices~[\citeauthor{pmid32575088}], therefore dramatically reduces the input size. However, it is less straightforward to perform such a separation under a Cone-Beam imaging geometry, as rays traverse multiple Cartesian planes. Motivated by the fact that a tomographic view only partially observes the desired volume during CBCT acquisition, we instead seek to reduce computation by including only the observed volume when synthesizing a tomographic view.

To this end, REGAS implements a Ray Path Transformation (RPT) operator $\mathcal{T}$. Given a view angle $a_k$, RPT transforms a CBCT volume from a Cartesian-based representation $V\in \mathbb{R}^{X\times Y\times Z}$ to view-based representation $R_{a_k}\in \mathbb{R}^{W\times H\times S}$, such that $P(a_k, w, h) = \sum_{s=1}^SR_{a_k}(w, h, s)$, where $S$ is the marching dimension of each ray, and $P(a_k) \in \mathbb{R}^{W\times H}$ is the tomographic view under angle $a_k$. Mathematically, the RPT operator $\mathcal{T}$ can be described as:
\begin{align}\label{eqn: RPT}
\begin{aligned}
    \mathcal{T}(V, a_k) = R_{a_k}, R_{a_k}(w,h,s) = V( \bm{l}^{a_k}_{w,h}(s)),
\end{aligned}
\end{align}
where $\bm{l}^{a_k}_{w,h}$ denotes the line segment that connects point $(w,h)$ in $P(a_k)$ and the ray source. As such, $R_{a_k}(w,h,s)$ is the minimum representation of $V$ to produce $P(a_k)$, and can be split and recombined over the $w-h$ plane while maintaining integrity to the Cone Beam forward projection geometry, as demonstrated in Fig.~\ref{fig:network}. This property enables patch training and for REGAS and scales REGAS to synthesize high resolution views. RPT is implemented through CUDA with gradient propagation, and is as fast as other GPU-based forward projection operations.

\subsection{View Synthesis Network (VSN)}

Two factors are essential in synthesizing reliable unobserved views for different phases. Firstly, the synthesized views of phase $i$ should be geometrically consistent with the observed views of the same phase, especially over the regions with large motion. Secondly, the synthesized views should be realistic and contain little noise in their texture. 

To ensure geometric consistency, we first obtain initial phase reconstructions $V^{\textrm{init}}_{i}=\mathcal{H}(P^{\textrm{obs}}_i)$, where $\mathcal{H}$ denotes a reconstruction method, e.g., FDK or SART. While $V^{\textrm{init}}_i$ have low imaging quality, they provide geometric templates to synthesize unobserved views. To synthesize views along angle $a_k$, we use RPT to obtain $R^{\textrm{init}}_{i,a_k}$ to reduce memory footprint. 

As shown in Fig.~\ref{fig:network}, the View Synthesis Network (VSN) consists of two subnets. The first subnet is a \underline{3D CNN $\mathcal{G}$} with a UNet-like architecture that removes artifacts in $R^{\textrm{init}}_{i,a_k}$. The forward process of $\mathcal{G}$ can be described as:
\begin{align}\label{eqn: 3DCNN}
\begin{aligned}
    \{R^{\textrm{ref}}_{1,a_k},R^{\textrm{ref}}_{2,a_k},...,R^{\textrm{ref}}_{N,a_k}\}=\mathcal{G}(R^{\textrm{init}}_{\forall i,a_k}) + R^{\textrm{init}}_{\forall i,a_k}\\
    R^{\textrm{init}}_{\forall i,a_k} = R^{\textrm{init}}_{1,a_k} \oplus R^{\textrm{init}}_{2,a_k}\oplus...\oplus R^{\textrm{init}}_{N,a_k},
\end{aligned}
\end{align}
where $\oplus$ concatenates $R^{\textrm{init}}_{i,a_k}$ over the channel dimension. The refined volumes $R^{\textrm{ref}}_{i,a_k}$ are constrained through a self-supervised loss, described as 
\begin{align}\label{eqn:3D_loss}
\mathcal{L}_{\textrm{MA}}=\lVert \frac{1}{N}\sum_{i}R^{\textrm{ref}}_{i,a_k} - R^{\textrm{ma}}_{a_k}\rVert_1, R^{\textrm{ma}}_{a_k}=\mathcal{T}(V^{\textrm{ma}}, a_k),
\end{align}
where $V^{\textrm{ma}}=\mathcal{H}(P^{\textrm{obs}})$ is the motion-affected reconstruction using all observed views. While $V^{\textrm{ma}}$ is blurry over moving anatomical regions, it is of high quality over static anatomical regions. Similarly, $R^{\textrm{ref}}_{i,a_k}$ on average should follow this pattern, espeically over empty space.

The second subnet is a \underline{2D CNN $\mathcal{F}$} based on RDN \cite{DBLP:conf/cvpr/ZhangTKZ018} that individually processes views from each phase and can be described as: \begin{align}\label{eqn: 2DCNN}
\begin{aligned}
    P^{\textrm{syn}}_{i}(a_k)=\mathcal{F}(P^{\textrm{ref}}_{i}(a_k))+P^{\textrm{ref}}_{i}(a_k),
\end{aligned}
\end{align}
where $P^{\textrm{ref}}_{i}(a_k) = \sum_{s}R^{\textrm{ref}}_{i}(a_k)$.
Residual learning is used in Eq. (\ref{eqn: 3DCNN}) and (\ref{eqn: 2DCNN}) to promote faster convergence.

Note that, for a given phase $i$, some of the synthesized views have corresponding $P^{\textrm{obs}}_{i}(a_k)$ during CBCT acquisition. We thus constraint its corresponding synthesis with an $\mathcal{L}_1$:

\begin{align}\label{eqn: recon_loss}
\mathcal{L}_{\textrm{rec}}=\lVert P^{\textrm{syn}}_i(a_k) - P^{\textrm{obs}}_i(a_k)\rVert_1.
\end{align}
Such a reconstruction loss also prevents Eq.~(\ref{eqn:3D_loss}) from making individual $R^{\textrm{ref}}_{i,a_k}$ the same as $R^{\textrm{ma}}_{a_k}$.

For those synthesized views without corresponding observations, an adversarial loss is employed to ensure that the synthesized views are statistically similar to the observed views of the same phase but at slightly rotated angles,
\begin{flalign}\label{eqn: 2DCNN_GAN}
\begin{aligned}
\mathcal{L}^{\mathcal{D}}_{\textrm{GAN}}=
\mathbb{E}&[\mathcal{D}(P^{\textrm{obs}}_i(a_k^*))-1)^2] +
\mathbb{E}[\mathcal{D}(P^{\textrm{syn}}_i(a_k))-0)^2],\\
\mathcal{L}^{\mathcal{G}}_{\textrm{GAN}}=\mathbb{E}&[\mathcal{D}(P^{\textrm{syn}}_i(a_k))-1)^2],
\end{aligned}
\end{flalign} 
where $P^{\textrm{obs}}_i(a_k^*)$ is a view from $P^{\textrm{obs}}_i$ and has the closet view angle $a_k^*$ to $a_k$. 


The overall loss for training the full REGAS is therefore:
\begin{align}\label{eqn:overall_loss}
\mathcal{L}_{\textrm{all}}=\lambda_{\textrm{MA}}\mathcal{L}_{\textrm{MA}} + \lambda_{\textrm{rec}}\mathcal{L}_{\textrm{rec}} + \lambda_{\textrm{GAN}}\mathcal{L}_{\textrm{GAN}}, 
\end{align}
where $\lambda_{\textrm{MA}}$, $\lambda_{\textrm{rec}}$, and  $\lambda_{\textrm{GAN}}$ are weights for their corresponding losses, and their values are empirically chosen as 1, 0.1, and 0.05, respectively.





\begin{table*}[!htb]
\setlength{\tabcolsep}{1pt}
\centering
\begin{tabular}{| l|l|l|l|l|l|l|l|l|l|l|l|}
\hline
 Phase & 1 & 2 & 3 & 4 & 5 & 6 & 7 & 8 & 9& 10 &Average\\
\hline

OSSART 
& 37.4/.962
& 36.8/.956
& 34.3/.917
& 36.6/.953
& 36.8/.956
& 34.8/.923
& 34.0/.910
& 33.0/.889
& 34.8/.924
& 34.5/.918
& 35.3/.931
\\

U-Net
&  37.4/.966
&  37.4/.967 
&  35.7/.953
&  37.3/.967
&  37.4/.967
&  36.3/.959
&  35.4/.951
&  34.7/.943
&  36.0/.955
&  36.0/.956
&  36.4/.959
\\

DL-PICCS
&  39.0/.972
&  38.9/.973
&  36.9/.959
&  38.8/.972
&  39.0/.973
&  37.5/.965
&  36.5/.953
&  35.5/.945
&  37.0/.956
&  37.3/.962
&  37.6/.963
\\

$\textrm{REGAS}^{\textrm{noDVF}}$
&40.6/.980
&40.2/.978
&38.8/.970
&40.0/.977
&40.1/.977
&39.1/.974
&37.9/.961
&36.8/.951
&38.5/.966
&38.5/.966
&39.1/.970
\\

\hline
\hline

$\textrm{OSSART}^{\textrm{TTV}}$ 
&  38.3/.975
&  38.2/.975
&  37.6/.971
&  38.0/.974
&  38.1/.974
&  37.2/.968
&  36.7/.962
&  36.4/.958
&  36.5/.959
&  36.2/.954
&  37.3/.967
\\

SMEIR
&39.9/.977
&39.6/.977
&38.5/.973
&40.0/.978
&39.7/.977
&38.6/.973
&37.5/.968
&36.0/.958
&38.3/.971
&38.2/.971
&38.6/.972
\\

REGAS

&\bf{43.6/.989}
&\bf{43.4/.988}
&\bf{42.8/.987}
&\bf{43.6/.989}
&\bf{43.4/.988}
&\bf{42.4/.986}
&\bf{41.8/.984}
&\bf{41.2/.982}
&\bf{42.4/.986}
&\bf{42.1/.985}
&\bf{42.7/.986}
\\

\hline
DVF-GT

&43.8/.987
&43.9/.987
&43.3/.986
&44.1/.988
&43.8/.987
&42.9/.985
&42.5/.984
&41.9/.982
&43.1/.985
&43.0/.985
&43.3/.986
\\
\hline

\end{tabular}
\caption{Quantitative evaluation of 4D CBCT reconstruction on 4D Lung~\protect\cite{2017A} in terms of PSNR and SSIM. The best results are in {\bf bold} besides the reference DVF-GT. Phases that have more observed views have higher PSNRs in their OSSART reconstructions.}
\label{tab:sota_table}
\end{table*}
\subsection{Deformation-based Refinement}\label{sec:dvf}

Combining synthesized and observed views, we can reconstruct phases $V^{syn}_i=\mathcal{H}(\{P^{\textrm{obs}}_i,P^{\textrm{syn}}_i\})$ and obtain images with less severe sparse-view artifacts. However, the detail quality of $V^{syn}_i$ is still lacking due to the imperfect synthetic views. To address this, we apply a deformation-based refinement on $V^{syn}_i$. 
Specifically, we estimate the inter-phase Deformation Vector Fields (DVFs) $D_{i\rightarrow j}$ through VoxelMorph~\cite{DBLP:journals/tmi/BalakrishnanZSG19}, where $i,j$ indicate the moving and target phases. Based on $D_{i\rightarrow j}$, we perform the following optimization to refine $V^{\textrm{syn}}_i$:

\begin{align}\label{eqn:dvf_refine}
V^{\textrm{out}}_i=\argmin_{V^{\textrm{syn}}_i} \lVert P_j^{\textrm{obs}}(a_k) - A(k)D_{i\rightarrow j} V^{syn}_i\rVert_2, \forall \{k,j\},
\end{align}
where $A(k)$ is the projection matrix from angle $a_k$. This step significantly boosts the reconstruction quality as 1. synthetic views are not used in the optimization objective, and 2. the estimated $D_{i\rightarrow j}$ is of high quality due to the lack of artifacts in $V^{syn}_i$. In fact, we show by experiments that $D_{i\rightarrow j}$ is close to the upper-bound DVF estimation from groundtruth phases.


In summary, REGAS produces synthetic views to reduce artifacts in $V^{syn}_i$ in a self-supervised manner; It then leverages $V^{syn}_i$ to obtain reliable DVFs such that we can update the reconstructions from only acquired views. As such, REGAS allows the reconstructions to satisfy multi-view consistency from all angles without constraining them to synthetic data. 
While deformation-based refinement is not restricted to $V^{syn}$ and can be applied to $V^{init}$, we find that $V^{syn}$ is a much better initialization and leads to better results, as the optimization process in Eq.~(\ref{eqn:dvf_refine}) is non-convex.

\section{Experiments}
\textbf{Dataset.} To quantitatively evaluate the effectiveness of respiratory-gated CBCT reconstruction, we follow previous works~\cite{wang2013simultaneous,pmid32575088,chee2019mcsart} in simulating the acquisition process based on Digitally Reconstructed Radiographs (DRRs)~\cite{milickovic2000ct}. Specifically, we use the 4D Lung dataset~\cite{2017A}, which contains 62 high quality ten-phased 4D Fan-Beam CTs, and the projection geometry from real respiratory-gated acquisitions~\cite{shieh2019spare}. We set detector size to $\{W=256,H=192\}$, detector spacing to $1.55mm$, and the total number of acquired views to $K=680$. CT volumes are normalized to an isotropic resolution of $1.5mm$ and a resolution of $128\times256\times256$. While REGAS can be scaled to higher resolutions than $P \in \mathbb{R}^{256\times 192}$ and $V \in \mathbb{R}^{128\times 256\times 256}$, these resolutions are selected to ensure that other comparative methods stay computationally feasible in memory and time. The phase-view correspondences are based on real measurements in~\cite{shieh2019spare} after aligning the max-inhale and max-exhale phases with~\cite{2017A}. We use OSSART to obtain $V^{\textrm{init}}_{i}$. For RPT, the ray marching dimension is set to $S=128$; during training $R^{\textrm{init}}_{i,a_k}$ are separated into patches of $128\times128\times128$. While our REGAS method is self-supervised, thus requiring no training data, the comparative DL-based supervised methods (U-Net and DL-PICCS) are trained on LIDC-IDRI~\cite{armato2011lung}, which contains 890 FBCT lung images after pre-processing. The same data generation method is used to obtain the undersampled reconstructions as inputs, where for each image three random phases and their respective $P^{obs}_i$ are selected.

\subsection{Implementation Details}

The VSN in REGAS is trained end-to-end with an Adam optimizer, which uses a momentum of 0.5 and a learning rate of 0.0001 until convergence. For 3D CNN, REGAS uses a 3D-UNet \cite{DBLP:conf/miccai/RonnebergerFB15} structure, where five downsampling and upsampling CNN layers are used. A 2D Residual DenseNet (RDN \cite{DBLP:conf/cvpr/ZhangTKZ018}) is used for the view-refining stage, where we use ten residual dense blocks, each of which has six convolution layers and growing rate of thirty-two. We use 10 NVIDIA 2080TI to train the network in one day. While patch training is used due to the memory cost of backpropagation, at inference time $P^{\textrm{syn}}$ are generated with full resolution. It takes 28.75ms to synthesize one view, or 176s to synthesize all views for one 4D CBCT image on a single NVIDIA 2080TI. We use VoxelMorph~\cite{DBLP:journals/tmi/BalakrishnanZSG19} for DVF estimation, where all $V^{\textrm{syn}}_i$ are used for training. We use 6 NVIDIA 2080TI GPUs to train VoxelMorph for one day to obtain the desired model. This is similarly done for all methods that use inter-phase modelling, i.e. $\textrm{SMEIR}^{\textrm{sim}}$ and DVF-GT. We use TIGRE\footnote{https://github.com/CERN/TIGRE} for OSSART reconstruction. Please see the supplemental material section for more details. 
\captionsetup[subfigure]{labelformat=empty}
\begin{figure*}[htb!]
    \setlength{\abovecaptionskip}{3pt}
    \setlength{\tabcolsep}{1pt}
    \centering
    \begin{tabular}[b]{|c|c|ccc|ccc|cc|}
        \hline
        Phase & OSSART & U-Net & DL-PICCS & $\textrm{REGAS}^{\textrm{noDVF}}$ & $\textrm{OSSART}^{\textrm{TTV}}$  & SMEIR & REGAS & DVF-GT &GT\\
        
        1 & 
        \begin{subfigure}[b]{0.100\linewidth}
            \includegraphics[width=\textwidth,height=\textwidth]{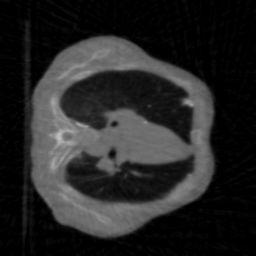}
        \end{subfigure} &
        \begin{subfigure}[b]{0.100\linewidth}
            \includegraphics[width=\textwidth,height=\textwidth]{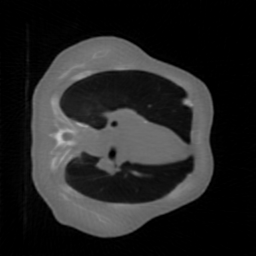}
        \end{subfigure} &
        \begin{subfigure}[b]{0.100\linewidth}
            \includegraphics[width=\textwidth,height=\textwidth]{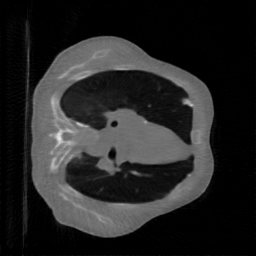}
        \end{subfigure} &
        \begin{subfigure}[b]{0.100\linewidth}
            \includegraphics[width=\textwidth,height=\textwidth]{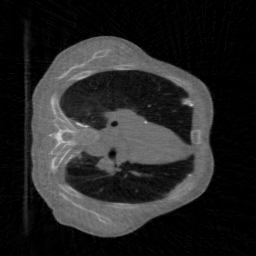}
        \end{subfigure} &
        \begin{subfigure}[b]{0.100\linewidth}
            \includegraphics[width=\textwidth,height=\textwidth]{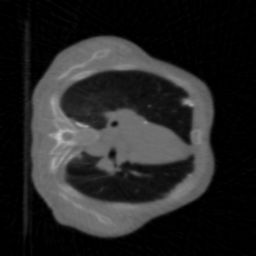}
        \end{subfigure} &
        \begin{subfigure}[b]{0.100\linewidth}
            \includegraphics[width=\textwidth,height=\textwidth]{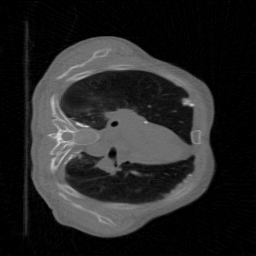}
        \end{subfigure} &
        \begin{subfigure}[b]{0.100\linewidth}
            \includegraphics[width=\textwidth,height=\textwidth]{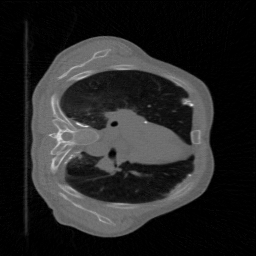}
        \end{subfigure} &
        \begin{subfigure}[b]{0.100\linewidth}
            \includegraphics[width=\textwidth,height=\textwidth]{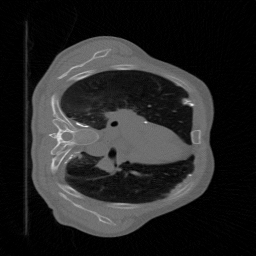}
        \end{subfigure} &
        \begin{subfigure}[b]{0.100\linewidth}
            \includegraphics[width=\textwidth,height=\textwidth]{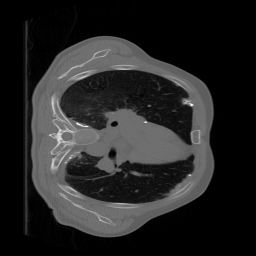}
        \end{subfigure} \\
         & 
        \begin{subfigure}[b]{0.100\linewidth}
            \includegraphics[width=\textwidth,height=\textwidth]{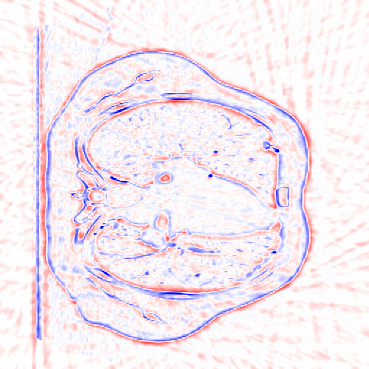}
            \caption{36.68/0.958}
        \end{subfigure} &
        \begin{subfigure}[b]{0.100\linewidth}
            \includegraphics[width=\textwidth,height=\textwidth]{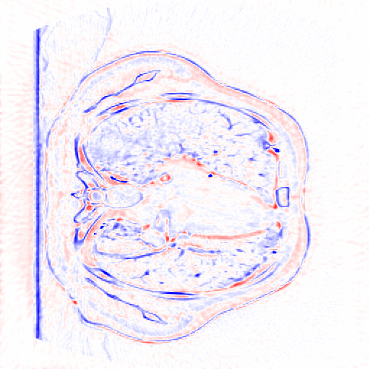}
            \caption{36.98/0.964}
        \end{subfigure} &
        \begin{subfigure}[b]{0.100\linewidth}
            \includegraphics[width=\textwidth,height=\textwidth]{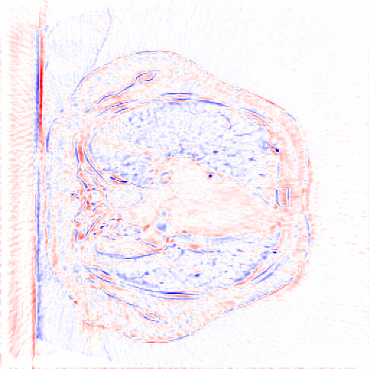}
            \caption{38.48/0.971}
        \end{subfigure} &
        \begin{subfigure}[b]{0.100\linewidth}
            \includegraphics[width=\textwidth,height=\textwidth]{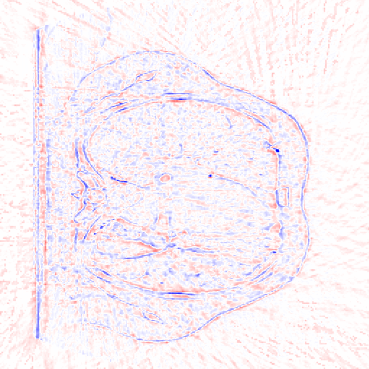}
            \caption{\underline{39.80/0.976}}
        \end{subfigure} &
        \begin{subfigure}[b]{0.100\linewidth}
            \includegraphics[width=\textwidth,height=\textwidth]{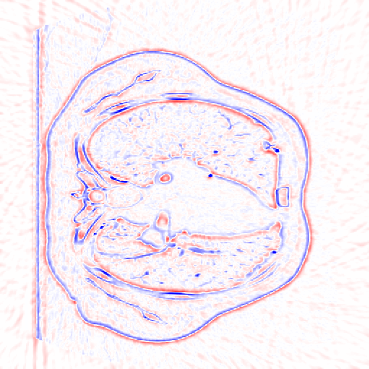}
            \caption{37.43/0.971}
        \end{subfigure} &
        \begin{subfigure}[b]{0.100\linewidth}
            \includegraphics[width=\textwidth,height=\textwidth]{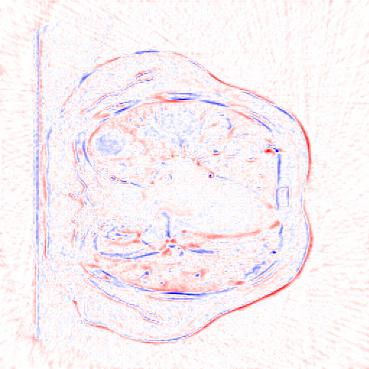}
            \caption{38.48/0.972}
        \end{subfigure} &
        \begin{subfigure}[b]{0.100\linewidth}
            \includegraphics[width=\textwidth,height=\textwidth]{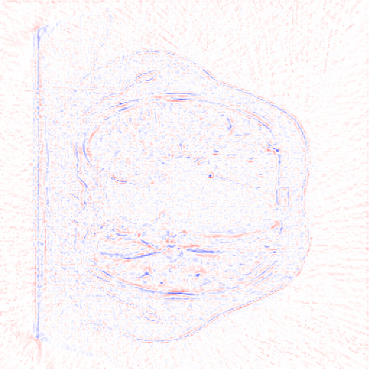}
            \caption{\bf{42.38/0.986}}
        \end{subfigure} &
        \begin{subfigure}[b]{0.100\linewidth}
            \includegraphics[width=\textwidth,height=\textwidth]{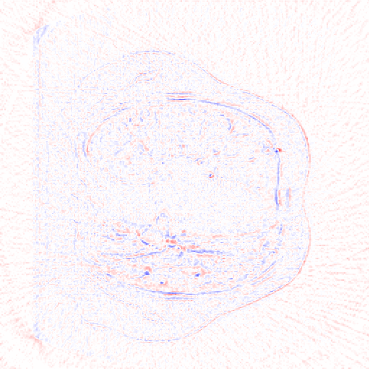}
            \caption{42.12/0.984}
        \end{subfigure} &
        \begin{subfigure}[b]{0.100\linewidth}
            \includegraphics[height=\textwidth]{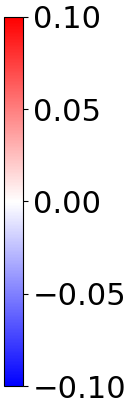}
            \caption{PSNR/SSIM}
        \end{subfigure} \\
        \hline
        
        5 & 
        \begin{subfigure}[b]{0.100\linewidth}
            \includegraphics[width=\textwidth,height=\textwidth]{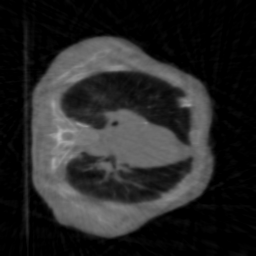}
        \end{subfigure} &
        \begin{subfigure}[b]{0.100\linewidth}
            \includegraphics[width=\textwidth,height=\textwidth]{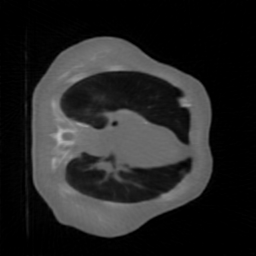}
        \end{subfigure} &
        \begin{subfigure}[b]{0.100\linewidth}
            \includegraphics[width=\textwidth,height=\textwidth]{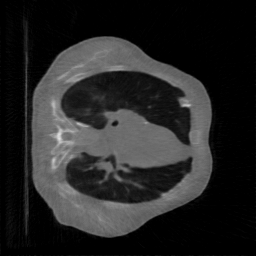}
        \end{subfigure} &
        \begin{subfigure}[b]{0.100\linewidth}
            \includegraphics[width=\textwidth,height=\textwidth]{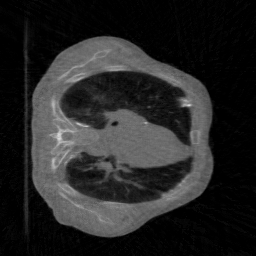}
        \end{subfigure} &
        \begin{subfigure}[b]{0.100\linewidth}
            \includegraphics[width=\textwidth,height=\textwidth]{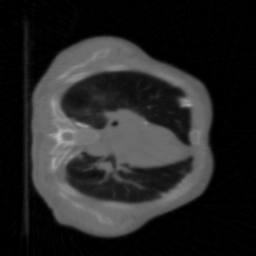}
        \end{subfigure} &
        \begin{subfigure}[b]{0.100\linewidth}
            \includegraphics[width=\textwidth,height=\textwidth]{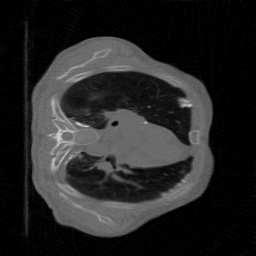}
        \end{subfigure} &
        \begin{subfigure}[b]{0.100\linewidth}
            \includegraphics[width=\textwidth,height=\textwidth]{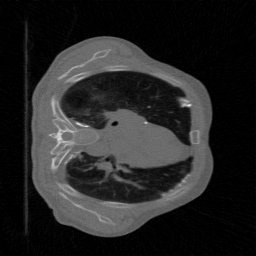}
        \end{subfigure} &
        \begin{subfigure}[b]{0.100\linewidth}
            \includegraphics[width=\textwidth,height=\textwidth]{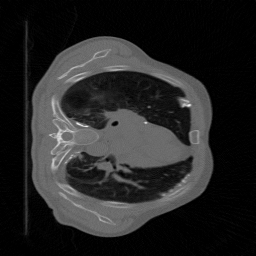}
        \end{subfigure} &
        \begin{subfigure}[b]{0.100\linewidth}
            \includegraphics[width=\textwidth,height=\textwidth]{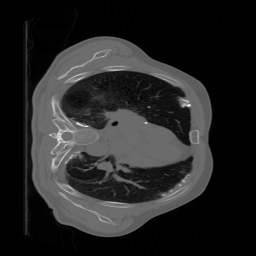}
        \end{subfigure} \\
         & 
        \begin{subfigure}[b]{0.100\linewidth}
            \includegraphics[width=\textwidth,height=\textwidth]{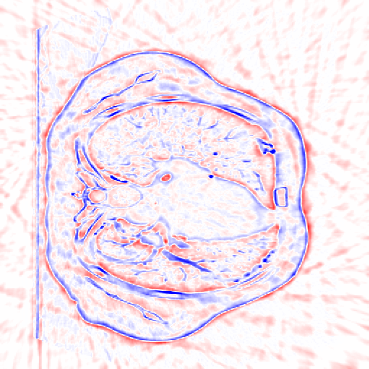}
            \caption{36.12/0.952}
        \end{subfigure} &
        \begin{subfigure}[b]{0.100\linewidth}
            \includegraphics[width=\textwidth,height=\textwidth]{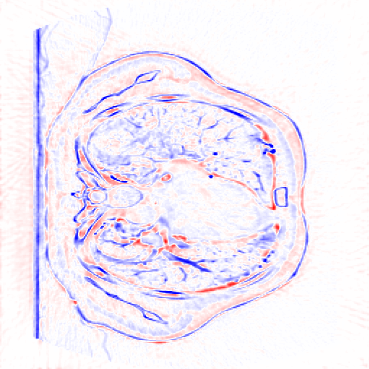}
            \caption{36.75/0.963}
        \end{subfigure} &
        \begin{subfigure}[b]{0.100\linewidth}
            \includegraphics[width=\textwidth,height=\textwidth]{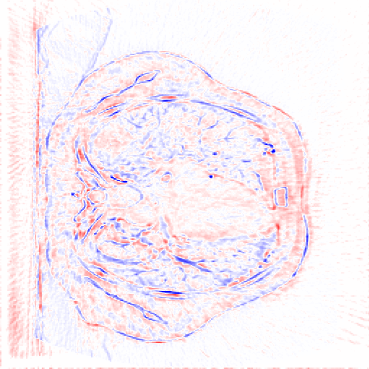}
            \caption{38.42/0.971}
        \end{subfigure} &
        \begin{subfigure}[b]{0.100\linewidth}
            \includegraphics[width=\textwidth,height=\textwidth]{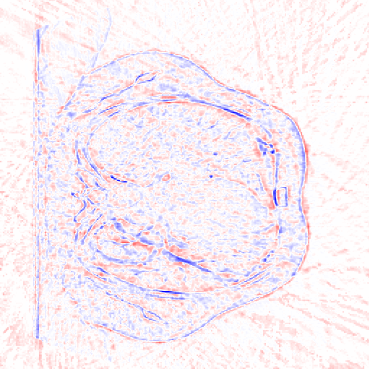}
            \caption{\underline{39.08/0.973}}
        \end{subfigure} &
        \begin{subfigure}[b]{0.100\linewidth}
            \includegraphics[width=\textwidth,height=\textwidth]{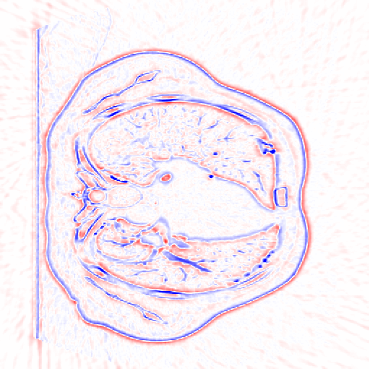}
            \caption{37.22/0.970}
        \end{subfigure} &
        \begin{subfigure}[b]{0.100\linewidth}
            \includegraphics[width=\textwidth,height=\textwidth]{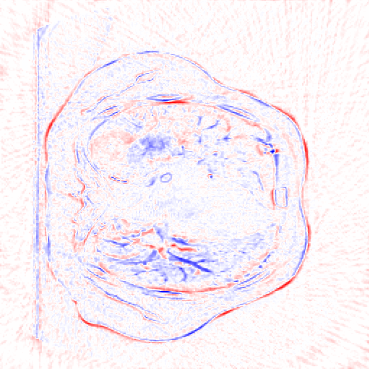}
            \caption{38.70/0.974}
        \end{subfigure} &
        \begin{subfigure}[b]{0.100\linewidth}
            \includegraphics[width=\textwidth,height=\textwidth]{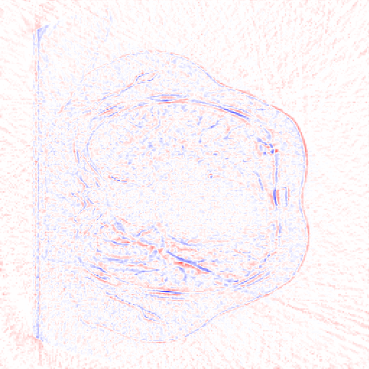}
            \caption{\bf{42.27/0.986}}
        \end{subfigure} &
        \begin{subfigure}[b]{0.100\linewidth}
            \includegraphics[width=\textwidth,height=\textwidth]{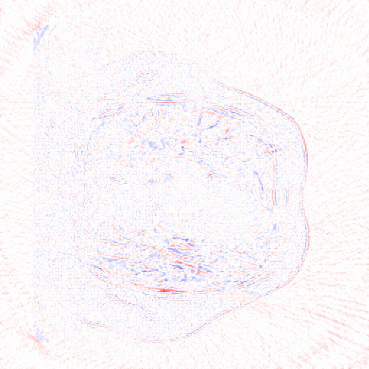}
            \caption{42.74/0.985}
        \end{subfigure} &
        \begin{subfigure}[b]{0.100\linewidth}
            \includegraphics[height=\textwidth]{imgs/diff_color/axis.png}
            \caption{PSNR/SSIM}
        \end{subfigure} \\
        \hline        
        
        \hline
        8 & 
        \begin{subfigure}[b]{0.100\linewidth}
            \includegraphics[width=\textwidth,height=\textwidth]{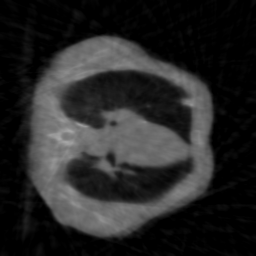}
        \end{subfigure} &
        \begin{subfigure}[b]{0.100\linewidth}
            \includegraphics[width=\textwidth,height=\textwidth]{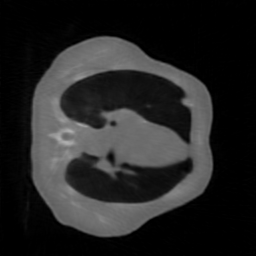}
        \end{subfigure} &
        \begin{subfigure}[b]{0.100\linewidth}
            \includegraphics[width=\textwidth,height=\textwidth]{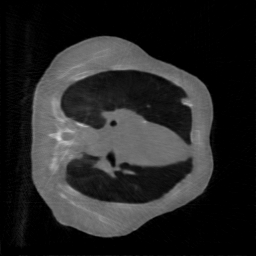}
        \end{subfigure} &
        \begin{subfigure}[b]{0.100\linewidth}
            \includegraphics[width=\textwidth,height=\textwidth]{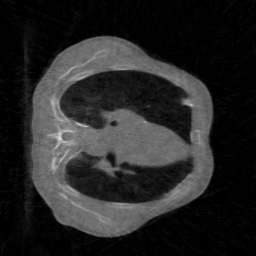}
        \end{subfigure} &
        \begin{subfigure}[b]{0.100\linewidth}
            \includegraphics[width=\textwidth,height=\textwidth]{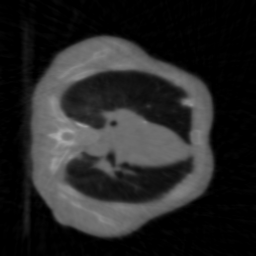}
        \end{subfigure} &
        \begin{subfigure}[b]{0.100\linewidth}
            \includegraphics[width=\textwidth,height=\textwidth]{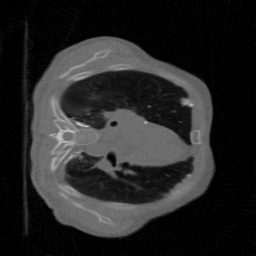}
        \end{subfigure} &
        \begin{subfigure}[b]{0.100\linewidth}
            \includegraphics[width=\textwidth,height=\textwidth]{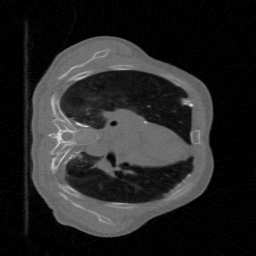}
        \end{subfigure} &
        \begin{subfigure}[b]{0.100\linewidth}
            \includegraphics[width=\textwidth,height=\textwidth]{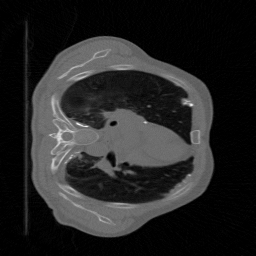}
        \end{subfigure} &
        \begin{subfigure}[b]{0.100\linewidth}
            \includegraphics[width=\textwidth,height=\textwidth]{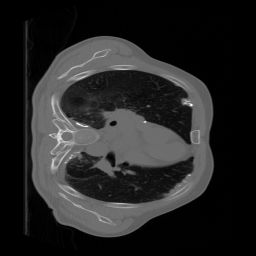}
        \end{subfigure} \\
         & 
        \begin{subfigure}[b]{0.100\linewidth}
            \includegraphics[width=\textwidth,height=\textwidth]{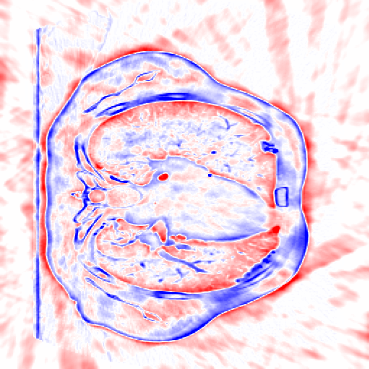}
            \caption{32.42/0.880}
        \end{subfigure} &
        \begin{subfigure}[b]{0.100\linewidth}
            \includegraphics[width=\textwidth,height=\textwidth]{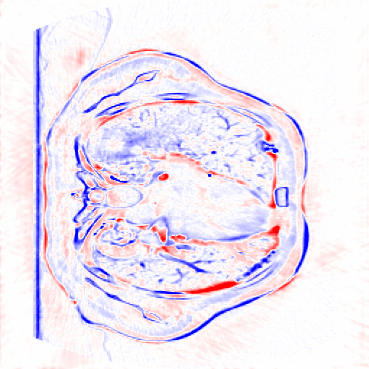}
            \caption{34.20/0.939}
        \end{subfigure} &
        \begin{subfigure}[b]{0.100\linewidth}
            \includegraphics[width=\textwidth,height=\textwidth]{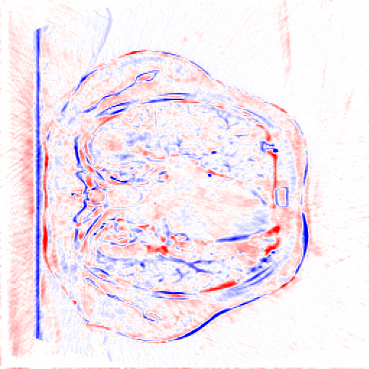}
            \caption{35.13/0.941}
        \end{subfigure} &
        \begin{subfigure}[b]{0.100\linewidth}
            \includegraphics[width=\textwidth,height=\textwidth]{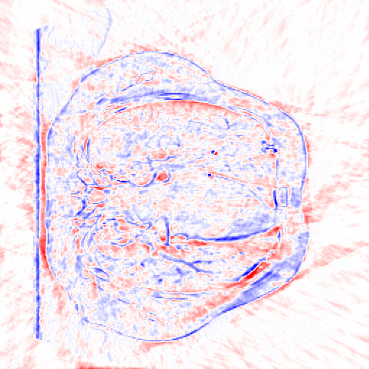}
            \caption{\underline{36.13/0.947}}
        \end{subfigure} &
        \begin{subfigure}[b]{0.100\linewidth}
            \includegraphics[width=\textwidth,height=\textwidth]{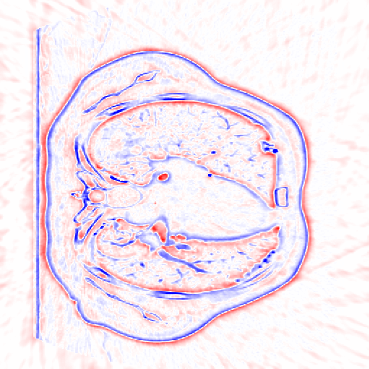}
            \caption{35.56/0.953}
        \end{subfigure} &
        \begin{subfigure}[b]{0.100\linewidth}
            \includegraphics[width=\textwidth,height=\textwidth]{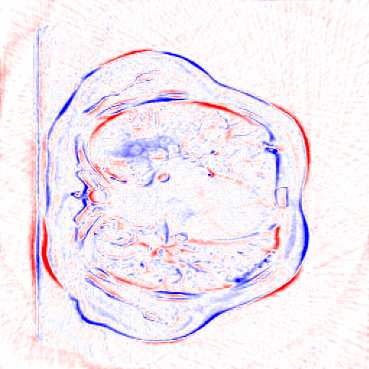}
            \caption{35.28/0.953}
        \end{subfigure} &
        \begin{subfigure}[b]{0.100\linewidth}
            \includegraphics[width=\textwidth,height=\textwidth]{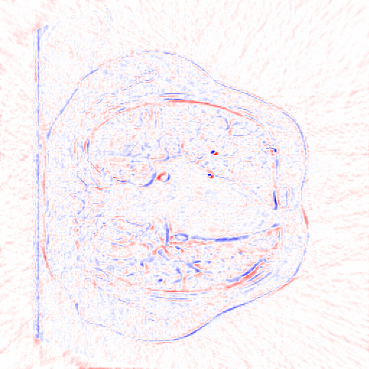}
            \caption{\bf{40.55/0.980}}
        \end{subfigure} &
        \begin{subfigure}[b]{0.100\linewidth}
            \includegraphics[width=\textwidth,height=\textwidth]{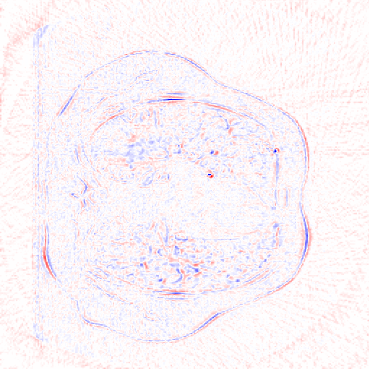}
            \caption{40.57/0.977}
        \end{subfigure} &
        \begin{subfigure}[b]{0.100\linewidth}
            \includegraphics[height=\textwidth]{imgs/diff_color/axis.png}
            \caption{PSNR/SSIM}
        \end{subfigure} \\        
        \hline
    \end{tabular}
    \caption{Visual comparisons of 4D reconstructions by different methods for phases 1 and 8. The best generation metrics are bold, the second best is underlined. Metrics are calculated volume-wise, and the difference maps are provided below the reconstruction images.}
    \label{tab:quant_img}
    \vspace{-1.5em}
\end{figure*}

\subsection{Quantitative Evaluations}
We summarize the quantitative comparisons of different reconstruction approaches against REGAS in Table \ref{tab:sota_table}. These approaches include:

\begin{itemize}
\item OSSART~\cite{andersen1984simultaneous}: A classic algebraic sparse-view reconstruction method; spatial TV is used to ensure smoothness in reconstructed images.
\item $\textrm{OSSART}^{\textrm{TTV}}$: OSSART with an additional Temporal TV term during reconstruction optimization, similar to~\cite{Ritschl_2012}.
\item U-Net: A supervised DL method based on the 3D U-Net architecture, representative of end-to-end approaches like~\cite{DBLP:journals/tmi/HanY18}.
\item DL-PICCS~\cite{https://doi.org/10.1002/mp.15183}: A multi-staged supervised DL method that applies PICCS between stages.
\item $\textrm{SMEIR}^{\textrm{sim}}$: A DVF-based optimization method based on~\cite{wang2013simultaneous}, where the DVFs are estimated from OSSART reconstructions.
\item DVF-GT: An upper-bound analysis on the DVF-based refinement approach, where the DVFs are obtained from groundtruth images. 
\end{itemize}

We visualize the results in Fig. \ref{tab:quant_img} over different phases. As we can see for OSSART, due to motion-induced irregular sampling, aliasing artifacts vary in phases 1 and 8, which have 105 and 50 $P^{\textrm{obs}}_i$ respectively. Details like rib bones are difficult to distinguish in phase 8, and clearer in phase 1. For clarity, we divide the rest of the baselines based on whether they model inter-phase information or not. 

\textbf{Single-phase-only modelling approaches.} Here we mainly focus on DL-based methods, as they are the current state of the art in sparse-view CT reconstruction without inter-phase modelling. The U-Net approach, which operates purely in the image space, clearly suffers from over-smoothing, especially in phase 8, as the pixel-wise loss leads to averaging when artifact is severe. This is somewhat improved in DL-PICCS, which contains two U-Nets. The results from the first U-Net are updated with $P^{\textrm{obs}}_i$ in an OSSART manner, which recovers some lost details that can be observed in OSSART. Despite such an update, the bone details are still not fully recovered in phase 8. 

We note that there are DL-based methods that attempt to leverage the temporal information, such as 4D-AirNet~\cite{pmid32575088}. However, 4D-AirNet is constrained to work on 2D slices instead of the native 3D space for Cone Beam geometry, as the GPU memory cost would be infeasible if the network is updated with 3D kernels. As such, it is restricted to Fan Beam geometry if view consistency is to be enforced. 

While DL-PICCS is chosen as a representative baseline, as it has a step to ensure view consistency like 4D-AirNet and is trained in two stages to avoid memory constraints; we note that DL-PICCS uses all memory in a NVIDIA 2080Ti GPU with a batch size of one.
Compute feasibility is an important factor in DL-based methods, and is a strength of REGAS. REGAS addresses the undersampling problem by synthesizing views instead of directly operating in the higher-dimension image space. With RPT, REGAS can be trained in patches while preserving CBCT geometry; this greatly scales its ability to use deeper networks while still generating high quality views. 
While RPT is used for reconstruction in REGAS, its functions to obtain minimum 3D representation with respect to a view and to propagate gradients are generally useful, such as for any DL-based method that looks to leverage CT and X-ray images together for learning. Finally, REGAS is self-supervised, and can avoid domain gaps such as scanner or patient differences from a training set. While $P^{\textrm{syn}}_i$ are not perfect and still lead to some artifacts, as in Fig.~\ref{tab:quant_img} for  $\textrm{REGAS}^{\textrm{noDVF}}$, bone details are much better recovered. 

\textbf{Inter-phase modelling approaches.} Here, the dominant approaches are non-DL-based approaches, which are less reliant on GPU memory and leverage the temporal correlations between phases. A common approach is to add a Temporal TV term during the reconstruction optimization process, as represented by $\textrm{OSSART}^{\textrm{TTV}}$. While $\textrm{OSSART}^{\textrm{TTV}}$ greatly out-performs OSSART in PSNR, its superior performance is based on the fact that the phase voxels are mostly stationary with each other; however, details remain blurry due to movements, as such errors around the edges are very high. 

SMEIR focuses on finding the optimal DVFs between phases from limited acquisition, such that we can reconstruct phases with $P^{\textrm{obs}}$ as described in Eq.~(\ref{eqn:dvf_refine}). Specifically, it finds DVFs through an iterative scheme, where (\romannumeral 1) DVFs are first estimated from initial SART reconstructions, (\romannumeral 2) a reference phase is updated following Eq.~(\ref{eqn:dvf_refine}), and (\romannumeral 3) DVFs are updated based on observed views from other phases and the reference phase. Steps 2 and 3 are repeated to obtain better reference phase and DVFs. SMEIR contains many hyper-parameters for its iterative optimization process, and does not have a public implementation for reference. Furthermore, as stated by the authors, the DVFs are not guaranteed to converge. To avoid an unfaithful re-implementation, we simplify SMEIR to steps 1 and 2 with no iteration, referred to as $\textrm{SMEIR}^{\textrm{sim}}$. $\textrm{SMEIR}^{\textrm{sim}}$ is comparable to REGAS since both involve a reconstruction step and a DVF-based refinement step. Similarly, both should benefit from iterative updates if they converge. We observe that due to the superior reconstructions of REGAS from synthesized views, the DVFs are estimated with more accuracy and lead to high quality reconstructions. The artifacts from $\textrm{REGAS}^{\textrm{noDVF}}$ are gone, as only real acquisitions are used. In comparison, while the $\textrm{SMEIR}^{\textrm{sim}}$ results are seemingly alias-free, the erroneous DVFs lead to inconsistent deformations around the body surface, which can be observed in the phase 8 difference image in Fig.~\ref{tab:quant_img}.

\textbf{Upper-bound analysis.} We also perform an upper-bound analysis called DVF-GT, where the DVFs are estimated directly from groundtruth phases and used in Eq.~(\ref{eqn:dvf_refine}). Since there does not exist a perfect DVF that leads to lossless phase reconstruction, DVF-GT represents the upper bounds of the DVF-based refinement step. We find that REGAS approaches DVF-GT in performance, particularly in SSIM  where REGAS is on par with DVF-GT. As such, we believe that REGAS approaches the upper limit from DVF-based refinement.

\begin{figure}[!htb]
    \setlength{\abovecaptionskip}{3pt}
    \setlength{\tabcolsep}{1pt}
    \begin{tabular}[b]{cc}
        \begin{subfigure}[b]{.5\linewidth}
            \includegraphics[width=\textwidth]{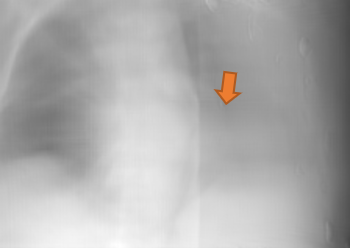}
            \caption{$\textrm{REGAS}^{\textrm{2D}}$}
        \end{subfigure} &
        \begin{subfigure}[b]{.5\linewidth}
            \includegraphics[width=\textwidth]{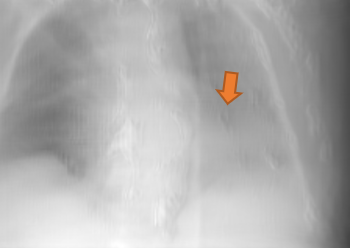}
            \caption{$\textrm{REGAS}^{\textrm{noRPT}}$}
        \end{subfigure} \\
        \begin{subfigure}[b]{.5\linewidth}
            \includegraphics[width=\textwidth]{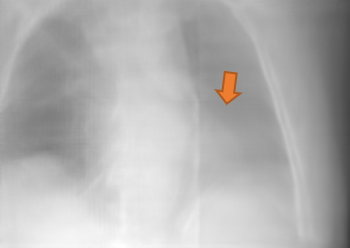}
            \caption{$\textrm{REGAS}^{\textrm{noMA}}$}
        \end{subfigure} &
        \begin{subfigure}[b]{.5\linewidth}
            \includegraphics[width=\textwidth]{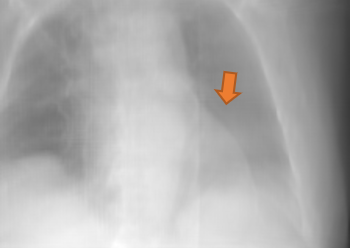}
            \caption{$\textrm{REGAS}^{\textrm{noDVF}}$}
        \end{subfigure} \\
    \end{tabular}
    \caption{Visual Comparison of synthesized projections over various ablated methods, taken from phase 7.}
    \label{fig:proj}
\end{figure}

\begin{table}[!htb]
    \setlength{\abovecaptionskip}{3pt}
    \setlength{\tabcolsep}{8pt}
    \centering
    \begin{tabular}{|c|c|c|c|c|}
    \hline
     Method & $\mathcal{G}$&RPT&$\mathcal{L}_{\textrm{MA}}$& Avg. Perf. \\
    \hline

    $\textrm{REGAS}^{\textrm{2D}}$ 
    & \ding{55}
    & \ding{55}
    & \ding{55}
    & 36.09/0.938
    \\
    \hline
    $\textrm{REGAS}^{\textrm{noRPT}}$
    & \ding{51}
    & \ding{55}
    & \ding{51}
    & 36.49/0.946
    \\
    \hline
    $\textrm{REGAS}^{\textrm{noMA}}$ 
    & \ding{51}
    & \ding{51}
    & \ding{55}
    & 37.96/0.964
    \\
    \hline
    $\textrm{REGAS}^{\textrm{noDVF}}$ 
    & \ding{51}
    & \ding{51}
    & \ding{51}
    & 39.05/0.970\\
    \hline
    




    
    \end{tabular}\\
    \caption{Ablation comparison among different implementations of REGAS. The average performance across all phases is presented.}
    \label{tab:ablation}
    \vspace{-1.3em}
\end{table}

\textbf{Ablation study.} We examine the effective of individual components in REGAS. $\textrm{REGAS}^{\textrm{2D}}$ trains the 2D CNN $\mathcal{F}$ with $\mathcal{L}_{\textrm{rec}}$ and $\mathcal{L}_{\textrm{GAN}}$ by directly using projections from $V^{\textrm{init}}_i$. With a 3D CNN $\mathcal{G}$, $\textrm{REGAS}^{\textrm{noRPT}}$ does not use RPT and has to downsample $V^{\textrm{init}}_i$ to $\mathbb{R}^{128\times 128\times 128}$ to be memory-wise viable. $\textrm{REGAS}^{\textrm{noMA}}$ does not use $\mathcal{L}_{\textrm{MA}}$, which denoises in 3D. As we can see in Table~\ref{tab:ablation}, adding $\mathcal{G}$, RPT, and $\mathcal{L}_{\textrm{MA}}$ each has a positive effect on performance. We show an instance of the synthesized views in Fig. \ref{fig:proj} for all versions of REGAS. In particular, we can see that views from $\textrm{REGAS}^{\textrm{noRPT}}$ and $\textrm{REGAS}^{\textrm{noMA}}$ both recover the rib cage frame on the right side of the lung. Compared to the full REGAS, there is less clarity over the heart region in $\textrm{REGAS}^{\textrm{noMA}}$ due to the noisier 3D CTs, and $\textrm{REGAS}^{\textrm{noRPT}}$ generates more fictitious details. The full REGAS produces a much more stable and geometrically consistent synthesis by leveraging DVF estimations from $\textrm{REGAS}^{\textrm{noRPT}}$ and real observations, which do not contain synthetic details. As such, we demonstrate the utility of each component and the effectiveness of scaling computation complexity with distributed forward projections. Please refer to the supplemental material on how volumetric resolution impacts views visually and other intermediate results.

\section{Conclusion}

We propose REGAS, a novel 4D CBCT reconstruction method from a single 3D CBCT acquisition. On a high level, REGAS first formulates sparse-view phase reconstruction as a synthesis problem on unobserved views. By combining observed and synthesized views, REGAS can obtain reconstructions with significantly less artifacts. REGAS then finds inter-phase DVFs based on these reconstructions, and use a synthesis-free optimization to refine reconstructions based on DVFs and observed views. Within REGAS, we propose RPT, which transforms a CT volume into a ray-based representation. The RPT-transformed volumes can be separated for distributed, patch-based training without affecting the Cone Beam imaging geometry, which addresses the memory bottleneck during network training. REGAS' network is trained in a self-supervised manner and does not require groundtruth on unobserved views, and is practical for clinical application. Our evaluation on a 4D Lung dataset show that REGAS significantly outperforms comparable methods in reconstruction performance. REGAS can assist in more accurate CBCT-based image-guided radiation therapy with high quality reconstructions. Since RPT reduces the memory footprint of 3D volume in a DL-based 3D-to-2D hybrid system, it can be applied to tasks that benefit from fusion learning on 3D CT and 2D radiographs, e.g. landmark detection, lesion segmentation, without sacrificing resolution.
\bibliography{main}


\end{document}


\appendix
\section{Implementation Details}
For 3D CNN, REGAS uses a 3D-UNet \cite{DBLP:conf/miccai/RonnebergerFB15} structure, where five downsampling and upsampling CNN layers are used. The feature channel sizes are $\{64,128,256,512,512\}$ for the downsampling layers, and $\{512,1024,512,256,10\}$ for the upsampling layers. A 2D Residual DenseNet (RDN \cite{DBLP:conf/cvpr/ZhangTKZ018}) is used for the view-refining stage, where we use ten residual dense blocks, each of which has six convolution layers and growing rate of thirty-two. We use VoxelMorph~\cite{DBLP:journals/tmi/BalakrishnanZSG19} for DVF estimation, where all $V^{\textrm{syn}}_i$ are used for training. We use 6 NVIDIA 2080TI GPUs to train VoxelMorph for one day to obtain the desired model. This is similarly done for all methods that use inter-phase modelling, i.e. $\textrm{SMEIR}^{\textrm{sim}}$ and DVF-GT. 
We use TIGRE\footnote{https://github.com/CERN/TIGRE} for OSSART reconstruction.

Ray Path Transformation (RPT) is implemented as a CUDA extension of PyTorch. Specifically, it follows cone beam geometry and uses nearest neighbor interpolation for ray marching in the discretely represented CT volumes. To avoid excessively long ray marching dimension, RPT only records rays when they are within the CT volume. Furthermore, the resolution of the ray marching dimension $s$ depends on the marching step size, i.e. if step size is very fine, the $s$ dimension in $R$ can be of very high resolution. As the $s$ dimension cannot be separated into patches due to preserving the integrity of forward projections, the high resolution potentially leads to memory issues. Empirically, we find that the a 256-step ray for a $128\times256\times256$ volume, i.e. $S=256$, can produce sufficiently accurate views; making finer step size yields very marginal improvements. We further propose a reasonable compromise to balance forward projection accuracy and the resolution in $s$ for higher dimension volumes. In particular, we can scale step size to any desired granularity. Since forward projection is a summation of ray marching values, we can downsample the $s$ dimension by performing a local summation on high resolution rays $\bm{l}_{HR}$, i.e. $\bm{l}_{LR}(x) = \sum_{i=0}^{k-1}\bm{l}_{HR}(kx+i)$ for $k$-time downsampling. As such the forward projections of $\bm{l}_{LR}$ and $\bm{l}_{HR}$ are equivalent, while the resulting $R$ still provides 3D context to a view as much as can be computationally handled. This is done in our experiments to bring the resolution of $s$ from $256$ to $128$. 

\begin{figure}[!htb]
    \setlength{\abovecaptionskip}{3pt}
    \setlength{\tabcolsep}{1pt}
    \begin{tabular}[b]{cc}
        \begin{subfigure}[b]{.5\linewidth}
            \includegraphics[width=\textwidth]{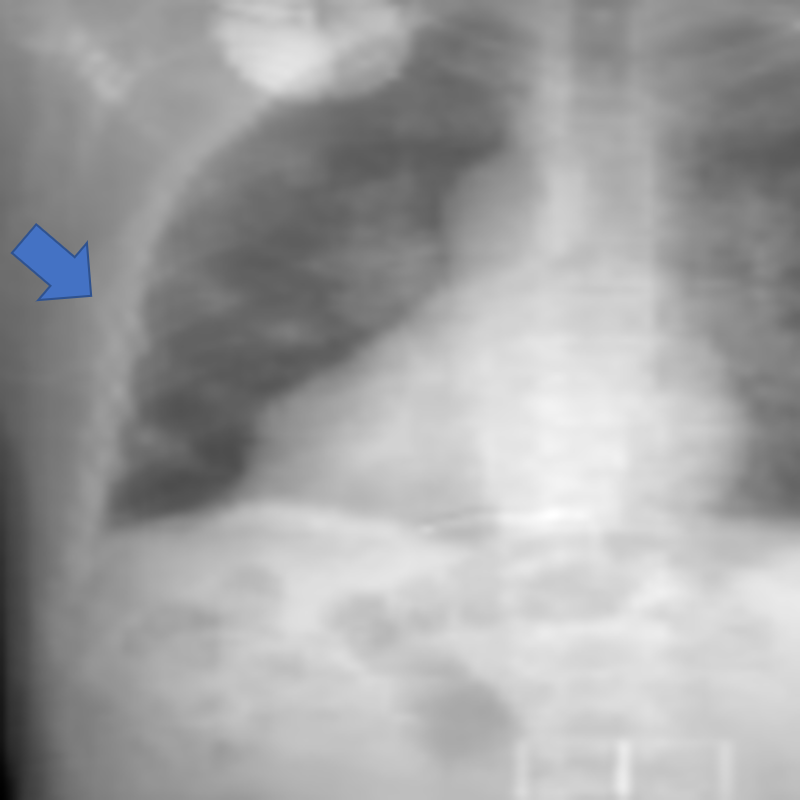}
            \caption{$\textrm{P}^{\textrm{LR}}$}
        \end{subfigure} &
        \begin{subfigure}[b]{.5\linewidth}
            \includegraphics[width=\textwidth]{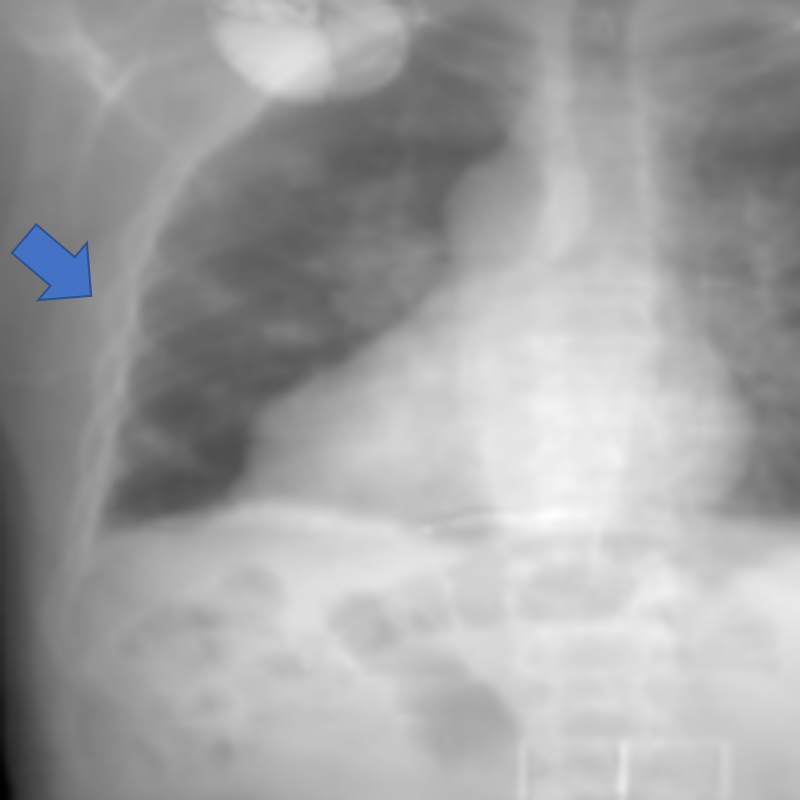}
            \caption{$\textrm{P}^{\textrm{HR}}$}
        \end{subfigure} \\
    \end{tabular}
    \caption{A visual comparison of forward projection quality. $P^{\textrm{HR}}$ and $P^{\textrm{LR}}$ are obtained from $V^{\textrm{HR}}\in \mathbb{R}^{128\times256\times256}$ and $V^{\textrm{LR}}\in \mathbb{R}^{128\times128\times128}$. The blue arrows indicate area where the bone details are lost in $P^{\textrm{LR}}$. }
    \label{fig:proj_res}
\end{figure}

Compared to alternative implementations which downsample the entire CT volume, RPT preserves high resolution information in a view-specific way. We demonstrate it in Fig. \ref{fig:proj_res}, where the same forward projection operation is done on a high resolution and low resolution CT $V^{\textrm{HR}}\in \mathbb{R}^{128\times256\times256}$, $V^{\textrm{LR}}\in \mathbb{R}^{128\times128\times128}$, yielding $P^{\textrm{HR}}$, $P^{\textrm{LR}}$ respectively. Note that $P^{\textrm{HR}}$ is what RPT produces. We can observe that small details are lost in $P^{\textrm{LR}}$, which also suffer from an overall blurriness.

\section{Data Processing and Simulation}
For simulation of 3D CBCT acquisitions from 4D FBCT, we follow the imaging parameters recorded in SPARE \cite{shieh2019spare}. Specifically, the input CT volumes have voxel spacings of $\{1.5, 1.5, 1.5\}$mm. The detector spacing is $\{1.55,1.55\}$mm and the detector dimension is $192\times256$, making the detector physical size $\{304.2,397.3\}$mm. The detector offset is $100$.

For pre-processing on the LIDC-IDRI \cite{armato2011lung} dataset, we discard all volumes that have a between-slices resolution higher than 2.5mm, and reshape the volumes into dimensions of $128\times256\times256$. The simulation of 3D CBCT acquisitions is performed similarly to that of \cite{2017A}. Specifically, for every CT in LIDC-IDRI, three initial phase reconstructions are simulated, where the phases are chosen at random. The training of U-Net and DL-PICCS transforms these OSSART-reconstructed phases to an artifact-free version through 3D-UNet structures \cite{DBLP:conf/miccai/RonnebergerFB15} using a $\mathcal{L}_1$ training loss. The feature channel sizes are of $\{48,96,192,384,384\}$ for five downsampling layers, and $\{384,768,384,192,1\}$ for five upsampling layers. The network size is smaller in comparison to the 3D-UNet in REGAS, as the input size is larger at $128\times256\times256$. Using batch size of one, this setup uses all memory available in a Nvidia 2080Ti. 

\captionsetup[subfigure]{labelformat=empty}
\begin{figure*}[htb!]
    \setlength{\abovecaptionskip}{3pt}
    \setlength{\tabcolsep}{1pt}
    \centering
    \begin{tabular}[b]{|cccccc|}
        \hline
        Phase & $\textrm{REGAS}^{\textrm{2D}}$ &  $\textrm{REGAS}^{\textrm{noRPT}}$ & $\textrm{REGAS}^{\textrm{noMA}}$ & $\textrm{REGAS}^{\textrm{noDVF}}$ & $\textrm{GT}$\\
        
        1 & 
        \begin{subfigure}[b]{0.190\linewidth}
            \includegraphics[width=\textwidth,height=\textwidth]{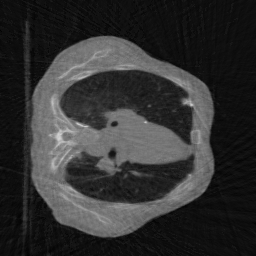}
            \caption{38.67/0.968}
        \end{subfigure} &
        \begin{subfigure}[b]{0.190\linewidth}
            \includegraphics[width=\textwidth,height=\textwidth]{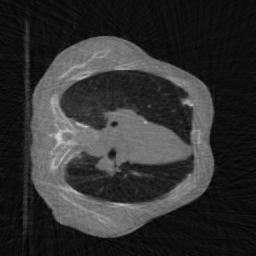}
            \caption{38.16/0.970}
        \end{subfigure} &
        \begin{subfigure}[b]{0.190\linewidth}
            \includegraphics[width=\textwidth,height=\textwidth]{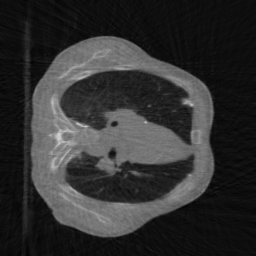}
            \caption{\underline{38.93/0.975}}
        \end{subfigure} &
        \begin{subfigure}[b]{0.190\linewidth}
            \includegraphics[width=\textwidth,height=\textwidth]{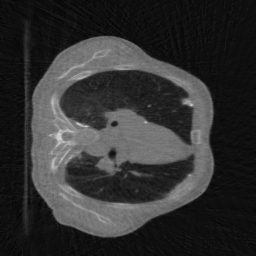}
            \caption{\bf{39.80/0.976}}
        \end{subfigure} &
        \begin{subfigure}[b]{0.190\linewidth}
            \includegraphics[width=\textwidth,height=\textwidth]{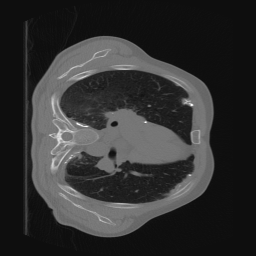}
            \caption{PSNR/SSIM}
        \end{subfigure} \\
        7 & 
        \begin{subfigure}[b]{0.190\linewidth}
            \includegraphics[width=\textwidth,height=\textwidth]{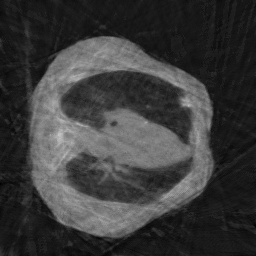}
            \caption{33.87/0.911}
        \end{subfigure} &
        \begin{subfigure}[b]{0.190\linewidth}
            \includegraphics[width=\textwidth,height=\textwidth]{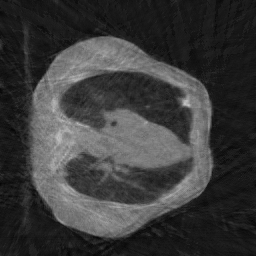}
            \caption{34.38/0.919}
        \end{subfigure} &
        \begin{subfigure}[b]{0.190\linewidth}
            \includegraphics[width=\textwidth,height=\textwidth]{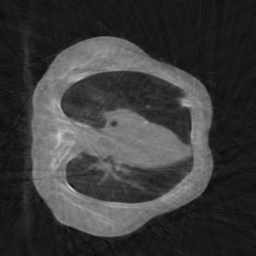}
            \caption{\underline{36.09/0.949}}
        \end{subfigure} &
        \begin{subfigure}[b]{0.190\linewidth}
            \includegraphics[width=\textwidth,height=\textwidth]{imgs/imags_ablation/100_HM10395_07-02-2003-p4-14571_0_recon_ours.png}
            \caption{\bf{36.87/0.954}}
        \end{subfigure} &
        \begin{subfigure}[b]{0.190\linewidth}
            \includegraphics[width=\textwidth,height=\textwidth]{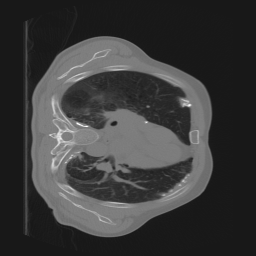}
            \caption{PSNR/SSIM}
        \end{subfigure} \\
        \hline
    \end{tabular}
    \caption{Visual comparisons of 4D reconstructions from ablation study. The best generation metrics are bold, the second best is underlined. Metrics are calculated volume-wise.}
    \label{tab:ablation_img}
\end{figure*}
\section{Ablation Study}

We examine the effectiveness of RPT and each component within the network structure by experimenting alternative implementations, including:

\begin{itemize}
\item $\textrm{REGAS}^{\textrm{2D}}$: We remove $\mathcal{G}$ of REGAS, and use only $\mathcal{F}$ to improve the quality of unseen views directly from $V^{\textrm{init}}_{i}$; we use $\mathcal{L}_{\textrm{rec}}$ and $\mathcal{L}_{\textrm{GAN}}$ for training.
\item $\textrm{REGAS}^{\textrm{noMA}}$: We train REGAS without $\mathcal{L}_{\textrm{MA}}$.
\item $\textrm{REGAS}^{\textrm{noRPT}}$: We use REGAS without RPT, i.e. providing $V^{\textrm{init}}_{i}$ as inputs to $\mathcal{G}$ and perform forward projection on the output volume; we then use $\mathcal{F}$ to improve synthesized view quality. $V^{\textrm{init}}_{i}$ is bicubically downsampled to $128\times128\times128$ due to memory constraint. 
\end{itemize}

$\textrm{REGAS}^{\textrm{2D}}$ directly applies the 2D part of REGAS on views from $V^{\textrm{init}}_{i}$; however, these views can be very blurry due to undersampled reconstruction. Therefore, it is difficult for networks to recover views that are consistent with the underlying geometry. While $\textrm{REGAS}^{\textrm{noRPT}}$ performs better than $\textrm{REGAS}^{\textrm{2D}}$ both quantitatively and visually, its information loss due to downsampling in 3D leads to lost details in 2D views and requires $\mathcal{F}$ to recover them via GAN. $\textrm{REGAS}^{\textrm{noMA}}$ shows good performance compared to the other two ablated methods, but does not share patterns as well as the full REGAS in 3D due to the lack of $\mathcal{L}^{\textrm{MA}}$. We show an instance of the synthesized views in Fig \ref{fig:proj} for all versions of REGAS. In particular, we can see that views from $\textrm{REGAS}^{\textrm{noRPT}}$ and $\textrm{REGAS}^{\textrm{noMA}}$ both recover the rib cage frame on the right side of the lung. Compared to the full REGAS, there is less clarity over the heart region in $\textrm{REGAS}^{\textrm{noMA}}$ due to the noisier 3D CTs, and $\textrm{REGAS}^{\textrm{noRPT}}$ generates more fictitious details. The full REGAS produces a much more stable and geometrically consistent synthesis, thus demonstrating the utility of each component and the effectiveness of scaling computation complexity with distributed forward projections. Please refer to the supplemental material on how volumetric resolution impacts views visually and other intermediate results.

\section{Intermediate REGAS results}
\begin{figure}[!htb]
    \setlength{\abovecaptionskip}{3pt}
    \setlength{\tabcolsep}{1pt}
    \begin{tabular}[b]{ccc}
        \begin{subfigure}[b]{.33\linewidth}
            \includegraphics[width=\textwidth]{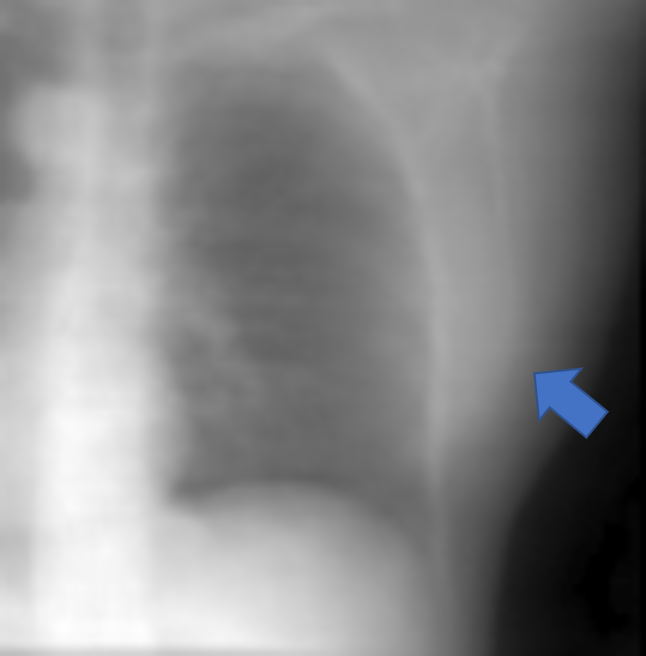}
            \caption{$\textrm{P}^{\textrm{init}}_{1,a_k}$}
        \end{subfigure} &
        \begin{subfigure}[b]{.33\linewidth}
            \includegraphics[width=\textwidth]{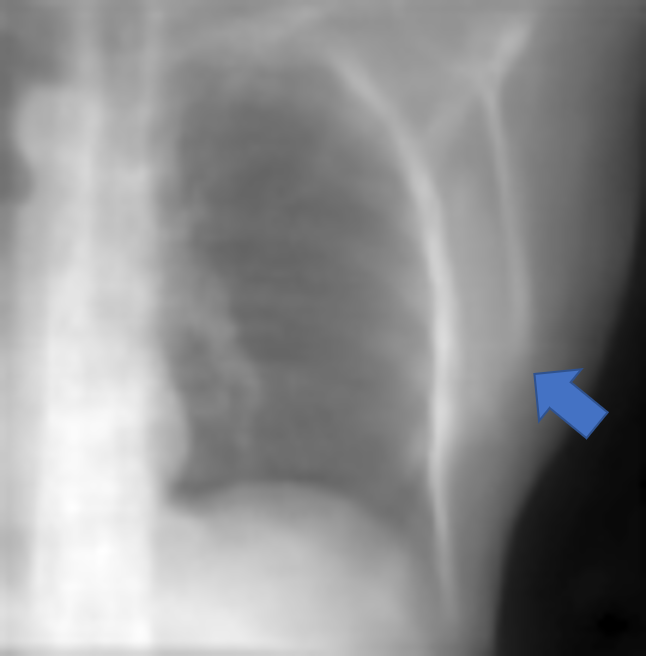}
            \caption{$\textrm{P}^{\textrm{ref}}_{1,a_k}$}
        \end{subfigure} &
        \begin{subfigure}[b]{.33\linewidth}
            \includegraphics[width=\textwidth]{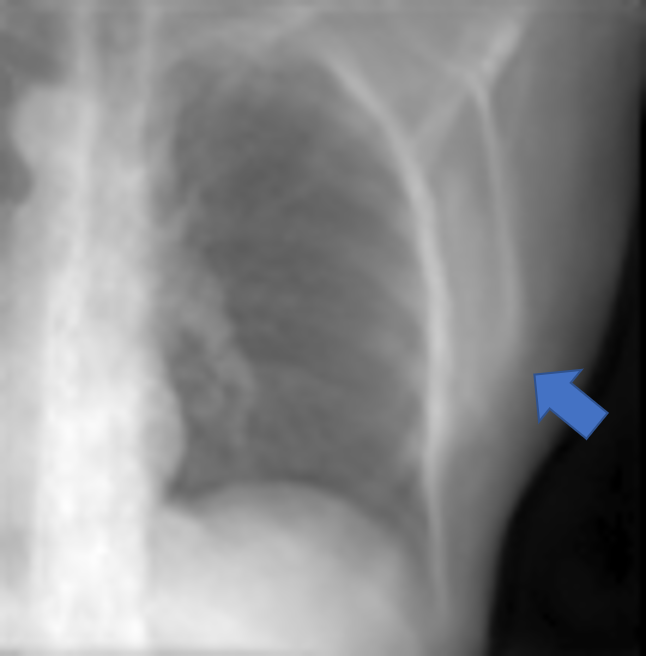}
            \caption{$\textrm{P}^{\textrm{syn}}_{1,a_k}$}
        \end{subfigure} \\
        \begin{subfigure}[b]{.33\linewidth}
            \includegraphics[width=\textwidth]{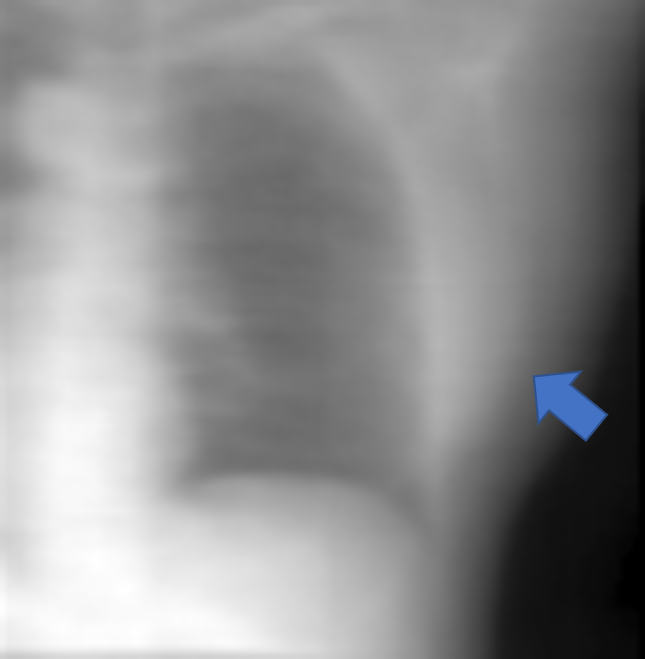}
            \caption{$\textrm{P}^{\textrm{init}}_{2,a_k}$}
        \end{subfigure} &
        \begin{subfigure}[b]{.33\linewidth}
            \includegraphics[width=\textwidth]{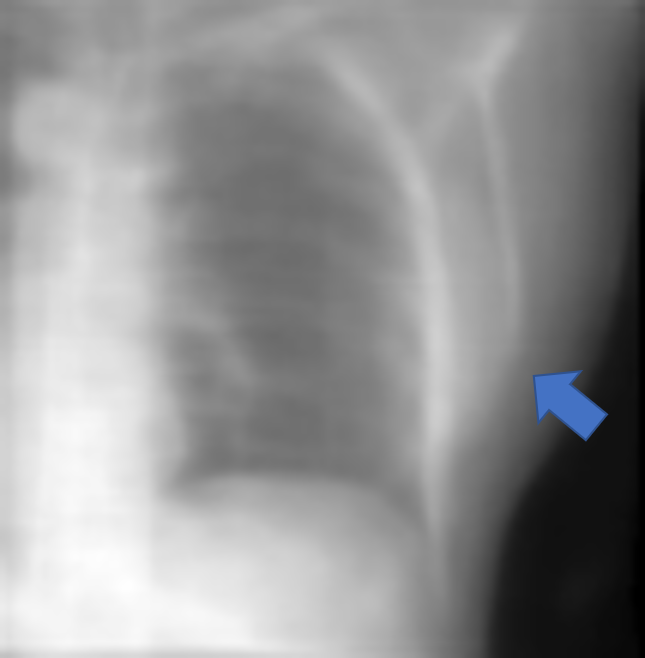}
            \caption{$\textrm{P}^{\textrm{ref}}_{2,a_k}$}
        \end{subfigure} &
        \begin{subfigure}[b]{.33\linewidth}
            \includegraphics[width=\textwidth]{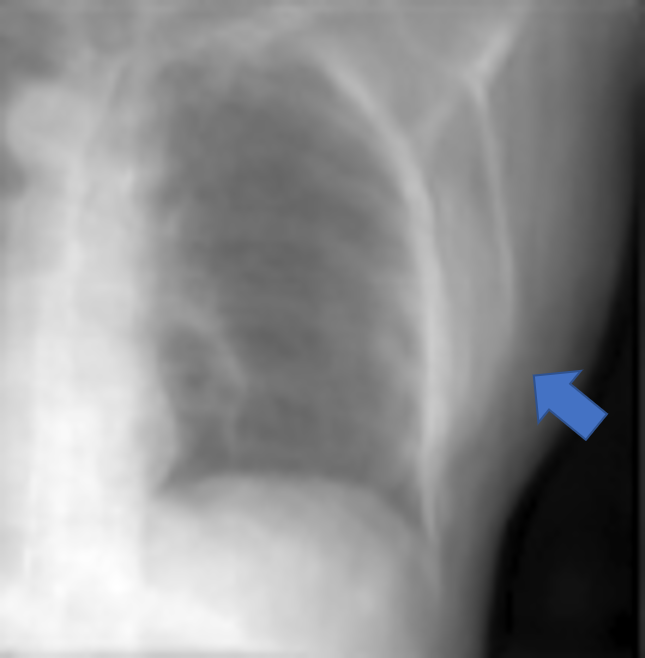}
            \caption{$\textrm{P}^{\textrm{syn}}_{2,a_k}$}
        \end{subfigure} \\
        \begin{subfigure}[b]{.33\linewidth}
            \includegraphics[width=\textwidth]{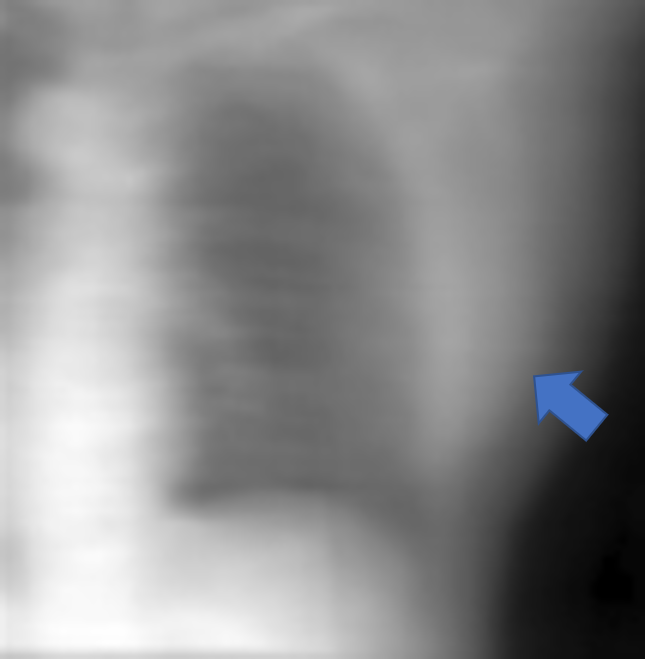}
            \caption{$\textrm{P}^{\textrm{init}}_{3,a_k}$}
        \end{subfigure} &
        \begin{subfigure}[b]{.33\linewidth}
            \includegraphics[width=\textwidth]{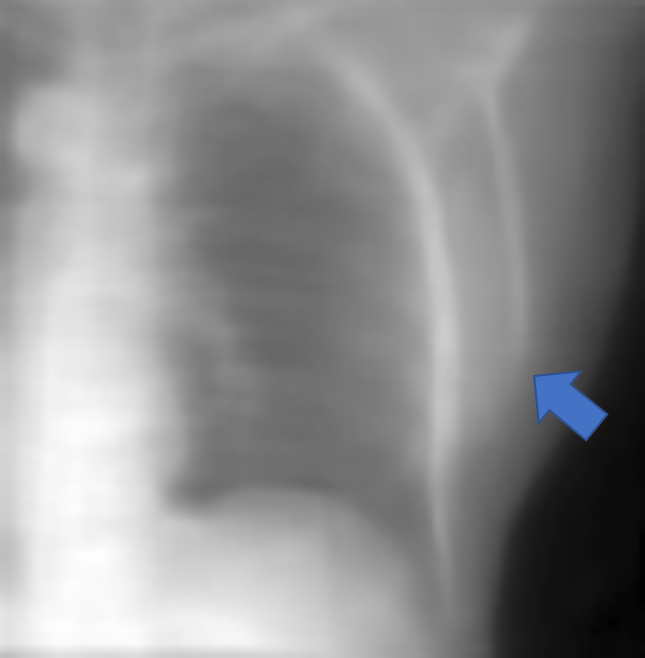}
            \caption{$\textrm{P}^{\textrm{ref}}_{3,a_k}$}
        \end{subfigure} &
        \begin{subfigure}[b]{.33\linewidth}
            \includegraphics[width=\textwidth]{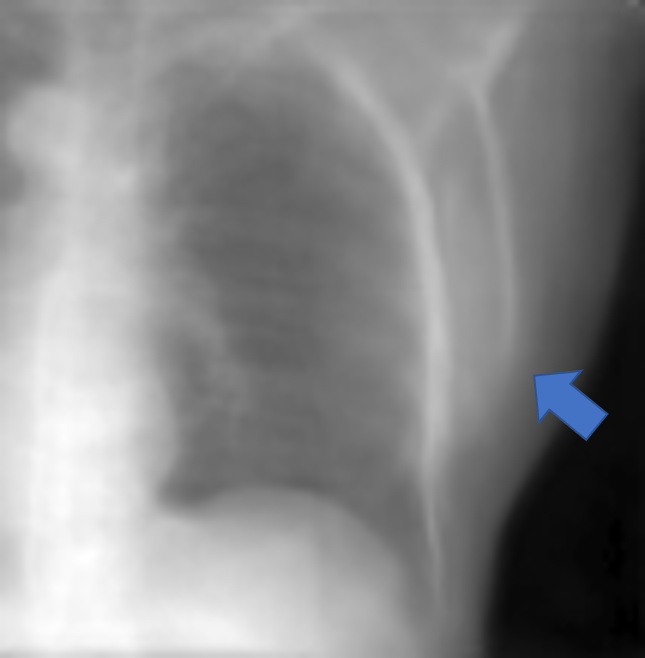}
            \caption{$\textrm{P}^{\textrm{syn}}_{3,a_k}$}
        \end{subfigure} \\
    \end{tabular}
    
    \begin{tabular}[b]{cc}
        \hspace{3.5em}
        \begin{subfigure}[b]{.33\linewidth}
            \includegraphics[width=\textwidth]{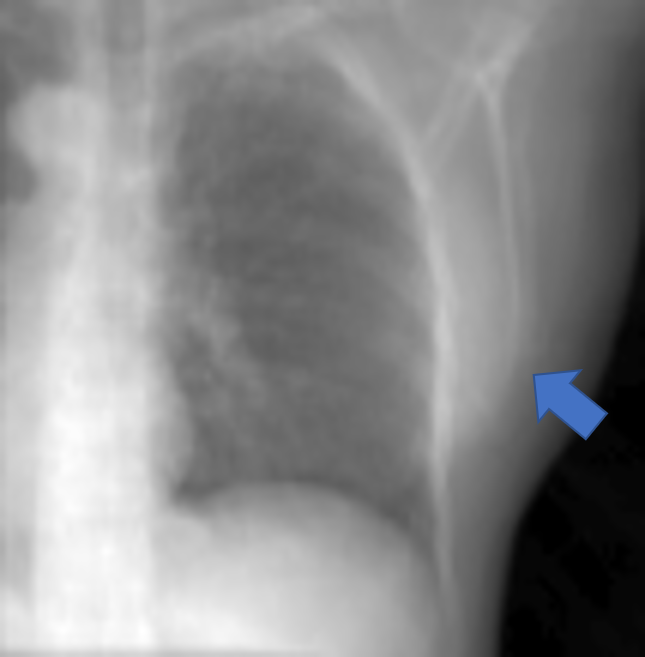}
            \caption{$\textrm{P}^{\textrm{obs}}_{1,a_k}$}
        \end{subfigure} &
        \begin{subfigure}[b]{.33\linewidth}
            \includegraphics[width=\textwidth]{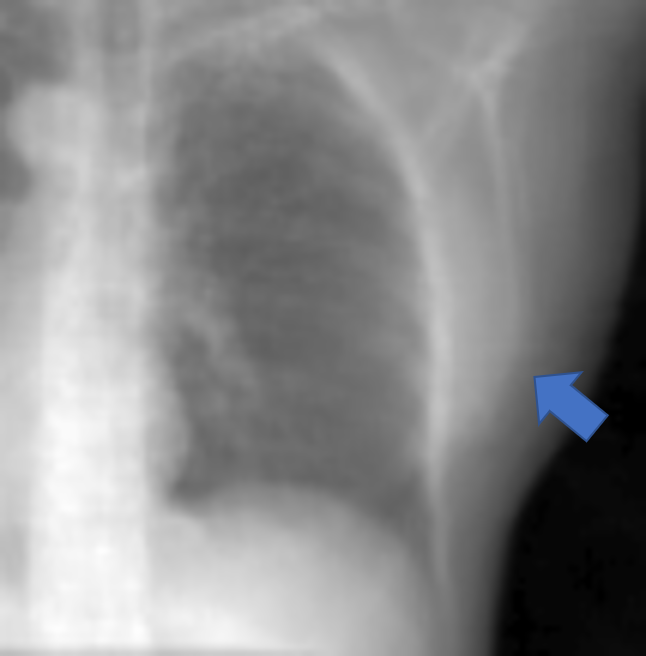}
            \caption{$\textrm{P}^{\textrm{ma}}_{a_k}$}
        \end{subfigure} \\
    \end{tabular}
    \caption{REGAS's intermediate results for phases $i=\{1,2,3\}$ are visualized as projections from view angle $a_k$. $\textrm{P}^{\textrm{init}}_{i,a_k}$,$\textrm{P}^{\textrm{ref}}_{i,a_k}$, and $\textrm{P}^{\textrm{syn}}_{i,a_k}$ are projections from the initial OSSART reconstructions, after going through $\mathcal{G}$, and after going through $\mathcal{F}$. The observed projection $\textrm{P}^{\textrm{obs}}_{1,a_k}$ and projection from $V^{ma}$ are also shown for reference.}
    \label{fig:projs_regas}
\end{figure}

In Fig. \ref{fig:projs_regas}, we show some intermediate results within REGAS that help visualize how image quality gradually improves through the various stages.

In the patient case shown in In Fig. \ref{fig:projs_regas}, $\textrm{P}^{\textrm{obs}}_{1,a_k}$ for phase one is observed in view angle $a_k$, while all other phases are not observed in this angle. As indicated by the blue arrows, we can see that the scapula is clearly visible in $\textrm{P}^{\textrm{obs}}_{1,a_k}$. However, that detail is covered by noise in the initial reconstructions of phase two and three, as shown in $\textrm{P}^{\textrm{init}}_{2,a_k}$, $\textrm{P}^{\textrm{init}}_{3,a_k}$. In fact, the scapula is only remotely visible in $\textrm{P}^{\textrm{init}}_{1,a_k}$, which shows that OSSART reconstruction is unsatisfactory even in recovering observed views. After going through the 3D-CNN $\mathcal{G}$, the reconstructed phases contain significantly less noise. As such the scapula can be observed in $\textrm{P}^{\textrm{ref}}_{1,a_k}$, $\textrm{P}^{\textrm{ref}}_{2,a_k}$, and $\textrm{P}^{\textrm{ref}}_{3,a_k}$, indicating the utility of our proposed motion-averaging loss. After a GAN-based 2D-CNN $\mathcal{F}$, more details are enhanced in $\textrm{P}^{\textrm{out}}_{i,a_k}$. Not only does the scapula appear sharper, the ribs and the spine are also more defined. By using all $\textrm{P}^{\textrm{out}}_{i,a_k}$ and $\textrm{P}^{\textrm{obs}}_{1,a_k}$ to perform reconstruction, we can thus obtain a much cleaner and higher quality 4D CBCT reconstruction.
\section{Limitations}
While REGAS demonstrates excellent performances, it is based on DRRs. Performance on real 4D CBCT acquisition is difficult to be evaluated quantitatively, as there exists neither a large public dataset nor reliable groundtruth in real acquisitions. As a proof of concept study, REGAS follows all the mentioned baselines in its usage of simulated acquisitions. We note that the adversarial loss can be adopted to synthesize realistic CBCT views acquired under various settings, and will perform future study on clinical data. 
\section{Additional Visual Results}
For additional animated results and comparisons, please refer to the video attachment.

\bibliographystyle{named}
\bibliography{ijcai22}